\documentclass[pdflatex,sn-mathphys-num]{sn-jnl}% Math and Physical Sciences Numbered Reference Style

\usepackage{graphicx}%
\usepackage{multirow}%
\usepackage{amsmath,amssymb,amsfonts}%
\usepackage{amsthm}%
\usepackage{mathrsfs}%
\usepackage[title]{appendix}%
\usepackage{xcolor}%
\usepackage{textcomp}%
\usepackage{manyfoot}%
\usepackage{booktabs}%
\usepackage{algorithm}%
\usepackage{algorithmicx}%
\usepackage{algpseudocode}%
\usepackage{listings}%
\usepackage{graphicx}
\usepackage{calc}
\usepackage{mathtools}
\usepackage{xcolor}
\usepackage{subcaption}
\usepackage{tikz}
\usepackage{pgfplots}
\pgfplotsset{compat=1.10}
\pgfplotsset{colormap={bluewhitered}{color=(blue) color=(white) color=(red)}}
\pgfplotsset{every axis plot/.append style={line width=0.3mm}}
\usepgfplotslibrary{fillbetween}
\usetikzlibrary{pgfplots.groupplots,spy,calc}
\usetikzlibrary{matrix,shapes,arrows,positioning,chains,spy}

\def\centerarc[#1](#2)(#3:#4:#5)%
    { \draw[#1] ($(#2)+({#5*cos(#3)},{#5*sin(#3)})$) arc (#3:#4:#5); }

\tikzstyle{block} = [rectangle, draw, fill=white,
    text width=8em, text centered, rounded corners, minimum height=2em]

\tikzstyle{line} = [draw, -latex']

\pgfdeclareplotmark{filledbox}{%
  \path[fill]
    (-1\pgfplotmarksize,-1\pgfplotmarksize) to
    (1\pgfplotmarksize,-1\pgfplotmarksize) to
    (1\pgfplotmarksize,1\pgfplotmarksize) to
    (-1\pgfplotmarksize,1\pgfplotmarksize) to
    (-1\pgfplotmarksize,-1\pgfplotmarksize);
  \pgfusepathqfill% mark=* uses \pgfusepathqfillstroke instead
}

\usepackage{amsfonts}
\usepackage{amssymb}

\hypersetup{
  colorlinks=true,
  linkcolor=blue,
  citecolor=blue,
  urlcolor=blue
}
\usepackage[nameinlink,noabbrev,capitalise]{cleveref}  % This should almost certainly be loaded last

\colorlet{darkgrey}{white!20!black}
\colorlet{midgrey}{white!40!black}
\colorlet{lightgrey}{white!60!black}
\definecolor{red}{RGB}{207,18,61}
\definecolor{blue}{RGB}{0,137,189}
\definecolor{green}{RGB}{19,175,57}

\definecolor{RED}{RGB}{255,0,0}

\newcommand{\doublecite}[1]{\citeauthor{#1}~\cite{#1}}

\newcommand{\Rey}{\mbox{\textit{Re}}}
\newcommand{\Ca}{\mbox{\textit{Ca}}}

\usepackage{amsmath}

\newcommand{\norm}[1]{\left\lVert#1\right\rVert} % double-pipe norm
\makeatletter
\newcommand*{\diff}%
    {\@ifnextchar^{\DIfF}{\DIfF^{}}}
\def\DIfF^#1{%
    \mathop{\mathrm{\mathstrut d}}%
        \nolimits^{#1}\gobblespace}
\def\gobblespace{\futurelet\diffarg\opspace}
\def\opspace{%
    \let\DiffSpace\!%
    \ifx\diffarg(%
        \let\DiffSpace\relax
    \else
        \ifx\diffarg[%
            \let\DiffSpace\relax
        \else
            \ifx\diffarg\{%
                \let\DiffSpace\relax
            \fi\fi\fi\DiffSpace}
\makeatother

\theoremstyle{thmstyleone}%
\theoremstyle{thmstyletwo}%

\theoremstyle{thmstylethree}%

\begin{document}

\title[Controlling three-dimensional falling liquid films]{\Large \vspace{-2cm} Controlling three-dimensional falling liquid films: \\ from weighted-residual models to direct numerical simulation}

\author[1,2]{\fnm{Oscar A.} \sur{Holroyd}}\email{oscar@vanellus.tech}

\author[1]{\fnm{Susana N.} \sur{Gomes}}\email{Susana.Gomes@warwick.ac.uk}

\author*[1]{\fnm{Radu } \sur{Cimpeanu}}\email{Radu.Cimpeanu@warwick.ac.uk}

\affil[1]{\orgdiv{Mathematics Institute}, \orgname{University of Warwick}, \city{Coventry}, \postcode{CV4 7AL}, \country{UK}}

\affil[2]{\orgname{Vanellus Technologies Ltd.}, \city{Cambridge}, \postcode{CB1 2SN}, \country{UK}}

\abstract{
We consider the problem of controlling a three-dimensional falling liquid film using same-fluid blowing and suction through the substrate. Building upon recent results for the corresponding two-dimensional problem, we employ a reduced-dimensional weighted-residual model for the film height and downstream flux as the basis for controller design. Linear stability analysis is used to estimate the number of actuators required to stabilise the film, while a linear-quadratic regulator (LQR) framework is used to construct feedback controls for the linearised system.
The resulting gain matrix is then combined with observations from direct numerical simulation (DNS) of the three-dimensional Navier-Stokes equations, allowing controls derived from the reduced-dimensional model to be applied directly to the full flow. Through systematic numerical tests, we demonstrate that these controls successfully stabilise the flat-film solution across a physically relevant range of Reynolds numbers and domain configurations.
We further show that both the number and placement of actuators play a crucial role in control performance, with spanwise instabilities requiring additional actuator coverage in the cross-stream direction. Although the extension from two to three dimensions substantially increases the cost of controller construction, this cost is incurred entirely offline during the computation of the gain matrix, making the overall framework computationally tractable. 
Our results demonstrate that LQR feedback controls derived from a three-dimensional weighted-residual model retain predictive value when applied to fully resolved DNS, extending previous two-dimensional control methodologies and helping bridge the gap between theoretical control design and experimentally relevant falling-film flows.
}



\maketitle

\section{Introduction}
Thin liquid films flowing down inclined surfaces are a canonical fluid-mechanical system of both fundamental interest and practical importance, with many applications relying on understanding and controlling their interfacial dynamics.
In particular, the ability to accurately and robustly manipulate liquid film thicknesses is of central importance to technological areas such as coating, liquid-based cooling systems, glass production and high-precision manufacturing processes used in the production of screens, lenses and microchips. 
In many applications, such as coating and precision manufacturing, the primary objective is to maintain a defect-free flat interface. 
In other settings, particularly those involving heat transfer and cooling, deliberately driving the system towards highly corrugated interfaces can be advantageous due to the accompanying increase in surface area and transport rates.
The successful manipulation of liquid films therefore relies on a combination of mathematical modelling, numerical simulation and control-theoretic design. Over the past several decades, this has led to the development of a hierarchy of models and control strategies of varying complexity and fidelity.

From a theoretical standpoint, falling liquid films have given rise to a hierarchy of mathematical models spanning a wide range of complexity, from analytically tractable weakly nonlinear equations to reduced-dimensional long-wave models and, ultimately, direct numerical simulation (DNS) of the Navier-Stokes equations. Extensive reviews of these modelling approaches in the uncontrolled setting can be found in \cite{craster2009dynamics,kalliadasis2011falling}. This hierarchy has also provided the foundation for a broad range of control methodologies. Among these, same-fluid blowing and suction through point actuators in the substrate has emerged as one of the most widely studied actuation mechanisms, with the majority of existing work focusing on two-dimensional films possessing a one-dimensional interface. In this context, early investigations focused on the Kuramoto-Sivashinsky equation, a weakly nonlinear model for small perturbations of a flat interface. Feedback controls that stabilise a flat interface were initially proposed in \cite{armaou2000feedback}, with extensions to nontrivial steady state or travelling wave solutions developed in \cite{gomes2015controlling,gomes2017stabilizing}. The success of these controls motivated their generalisation to long-wave models that act as reduced-dimensional approximations of the interface, such as the Benney equation and the weighted residuals model in \cite{thompson2016stabilising}. While these results were promising, there is no guarantee these controls would be successful in an experimental setting; to this end, one can consider DNS of the multi-phase Navier-Stokes equations as a proxy to an experiment, referred to as \emph{in silico} experiments. The first attempt to use feedback controls to stabilise flat (and other shaped) interfaces in DNS was performed in \cite{cimpeanu2021active}, where proportional controls based on linear stability analysis of the Benney and weighted-residuals models were considered. There, purely proportional controls (of the form $f(x_i,t) = -\alpha (h(x_i-\delta,t)-1)$) were shown to successfully stabilise the flat interface. More recently, we developed an LQR framework for controlling falling films in DNS using gain matrices derived from weighted-residual models \cite{holroyd2024linear}. In addition to blowing and suction through the substrate, there is related work in controlling interfacial dynamics in several contexts, with examples including interfaces of electrified films, using the electric field as a control \cite{wray2022electrostatic} and combining a model predictive control methodology with DNS observations, explorations of using substrate heating as control~\cite{thompson2019Robust,pino2026JFM}, or controlling fluids with contact lines \cite{boujo2019pancake}.
    
Although the successful control of the two-dimensional film is an interesting result in and of itself, significant challenges must be addressed before the resulting controls can be applied in an experimental setting. One of these challenges is the issue of perfect, complete interfacial information---in an experimental setting, observations of the interface (and other quantities) may be limited in location or at some time intervals, and they are likely to be noisy. 
In \cite{holroyd2024linear}, this issue was investigated through partial and noisy observations of the interface. It was shown that stabilisation could still be achieved using a nonlinear estimator of the dynamics based on the weighted-residual model and limited observations of the interface only, suggesting a degree of robustness to realistic measurement limitations.
 
The second issue which must be addressed is the fact that real, physical fluid systems are three-dimensional. Even in scenarios such as Hele-Shaw cells~\cite{cuttle2023compression} or axisymmetric flows in narrow channels~\cite{dietze2015films}, any controls ultimately act on a three-dimensional physical system, even though two-dimensional models are often sufficiently accurate for prediction or analysis. Therefore, to close the gap between the existing control frameworks and experimentally realisable systems, we extend the hierarchy of models considered in previous work (e.g. \cite{thompson2016falling,cimpeanu2021active,holroyd2024linear}) to include an additional dimension in the plane of the flow. 

There is very little work on controlling falling liquid films in the full three-dimensional (3D) case, with an initial exploration in \cite{tomlin2019optimal} where the authors used the Kuramoto-Sivashinsky equation as an approximate model for the interface, showing
existence of an optimal (distributed) control capable of stabilising the flat solution and other desired states, and \cite{tomlin2019point}, which considered point-actuated control of the same model, showing feedback controls are effective when controlling the flat solution and other states, and exploring the effect of the number of actuators and their arrangement. More recently, \doublecite{pino2026integral} explored the modelling and control of a 3D liquid metal coating process, using reinforcement learning to design controls based on gas jets and electromagnetic actuators, successfully reducing the amplitude of instabilities in the films.

The aim of this paper is to demonstrate that feedback controls obtained from a linearised three-dimensional weighted-residual model remain effective when applied directly to DNS of the three-dimensional Navier-Stokes equations.  For this reason, we do not explore the control within the reduced-dimensional models, as done in \cite{thompson2016stabilising}, and instead apply controls directly to the \emph{in silico} experiments represented by the DNS. To our knowledge, this is the first demonstration of reduced-model-based feedback control of a three-dimensional falling film in fully resolved Navier-Stokes simulations. 
In \cref{sec:3d_models} we introduce the governing equations (conservation of momentum and mass) in three dimensions, and, using asymptotic methods, namely the weighted residual boundary layer methodology, we derive an accurate reduced-dimensional model which reduces our system to two partial differential equations (PDEs) for the height of the film and its downstream flux. Aiming to control the system towards a flat film, we then perform linear stability analysis to establish the number of unstable modes of the system in Section~\ref{sec:linstab}. 
This will in turn enable us to derive feedback controls using the Linear-quadratic regulator (LQR) methodology in Section ~\ref{sec:lqr_control}, where we also comment on the numerical complexity of solving the associated matrix equations. We finally present numerical results in Section~\ref{sec:3d-results}, where we demonstrate the efficiency of the derived controls and investigate how control performance depends on the Reynolds number, the domain size and aspect ratio, and the number and arrangement of point actuators. We conclude with a discussion on the implications of our results and ongoing work.

\section{The hierarchy of models in three dimensions}%
\begin{figure}[t]
    \centering
    \begin{tikzpicture}
      \input{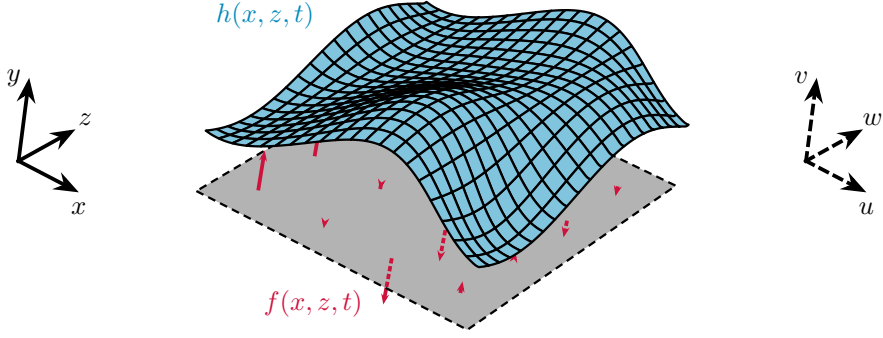}
    \end{tikzpicture}
    \caption{Diagram of the 3D falling liquid film control problem.
    }
    \label{fig:3d_diagram}
  \end{figure}

\label{sec:3d_models}
  We consider a 3D liquid film flowing down an inclined plane. Although the plane could, in principle, be tilted in an arbitrary horizontal direction, the governing dynamics are invariant under rotations about the vertical axis. We therefore, without loss of generality, choose coordinates such that the flow is directed along the $x$-axis, with the $z$-axis corresponding to the spanwise direction, and the $y$-axis measuring the distance normal to the substrate, as shown in \cref{fig:3d_diagram}.
  We impose periodic boundary conditions in both the streamwise ($x$-) and spanwise ($z$-) directions. While periodicity is an idealisation, it is a standard assumption in the analysis of thin-film flows, offering significant analytical and computational advantages. In particular, it permits a Fourier representation of the solution and facilitates the linear stability and control-theoretical analyses that follow. Similar assumptions have been adopted in studies of both the Kuramoto-Sivashinsky equation (e.g. \doublecite{kalogirou2015depth,tomlin2019point}) and weighted-residual thin-film models (e.g. \doublecite{ruyer2002further}). 

  Control is applied through point-actuated blowing and suction of the same fluid through the substrate. The resulting control input is represented by

\begin{equation}\label{eqn:control_general}
      f(x,z,t) = \sum_{i=1}^m d(x-x_i,z-z_i)a_i(t),
  \end{equation}
where $m$ denotes the number of actuators, $a_i(t)$ is the control amplitude associated with actuator $i$, and $(x_i,z_i)$ specifies its location on the substrate. The spatial distribution of each actuator is described by

 \begin{equation}
    \label{eqn:actuator-3d}d(x,z) = \alpha \exp \left( \frac{\cos(2\pi x / L_x) + \cos(2\pi z / L_z) - 2}{\omega^2} \right),
  \end{equation}
where $\omega$ controls the actuator width and $\alpha$ is a normalisation constant chosen such that $\int_0^{L_x}\int_0^{L_z} d(x,z) \diff x \diff z = 1$.
The control objective is then to determine the amplitudes $a_i(t)$ which, for a given number and arrangement of actuators, stabilise the Nusselt flat-film solution.

\subsection{The Navier-Stokes equations in 3D}\label{sec:NS}
The dynamics of the liquid film are governed by the three-dimensional incompressible Navier-Stokes equations, where the velocity vector is $\mathbf{u} = (u,v,w)$, with $u, \, v, \, w$ the streamwise, wall-normal, and spanwise velocities of the fluid, respectively, $p$ is the pressure, and $\theta$ is the inclination angle~\cite{kalliadasis2011falling}.
Let $h_s$ denote the mean film thickness. The uncontrolled system admits the uniform Nusselt solution $h(x,z,t)=h_s$~\cite{nusselt1923warm}, corresponding to a parabolic velocity profile with free-surface velocity $U_s = \frac{\rho g h_s^2 \sin \theta}{2\mu}$.
We nondimensionalise lengths using $h_s$, velocities using $U_s$, and pressures using the pressure scale $\frac{\mu U_s}{h_s}$, thus
defining the Reynolds and capillary numbers
\[
\Rey = \frac{\rho U_sh_s}{\mu}, \qquad \Ca = \frac{\mu U_s}{\gamma},
\]
which measure the relative importance of inertial and viscous terms, and of gravity and surface tension (represented by $\gamma$), respectively. 
Furthermore, we non-dimensionalise the $x$- and $z$ directions so that the domain we consider is $(x,z) \in [0,L_x]\times[0,L_z]$, the equations read
  \begin{align}
    \label{eqn:3d-momentum1}\Rey (u_t + uu_x + vu_y + wu_z) &= -p_x + 2 + u_{xx} + u_{yy} + u_{zz},\\
    \label{eqn:3d-momentum2}\Rey (v_t + uv_x + vv_y + wv_z) &= -p_y - 2\cot\theta + v_{xx} + v_{yy} + v_{zz}, \\
    \label{eqn:3d-momentum3}\Rey (w_t + uw_x + vw_y + ww_z) &= -p_z + w_{xx} + w_{yy} + w_{zz}, \\
    \label{eqn:3d-continuity}u_x + v_y + w_z &= 0.
  \end{align}

Note that \cref{eqn:3d-momentum3} contains no gravitational forcing terms. This is because the coordinate system is chosen such that gravity acts entirely within the $xy$-plane and therefore has no component in the spanwise direction. The free-surface boundary conditions (at \(y=h(x,z,t)\)) are:
  \begin{align}
    \begin{split}
      \label{eqn:3d-stress1}p &- \frac{2(u_x+w_z)h_xh_z + h_x^2u_x + h_z^2w_z - (u_y+v_x)h_x - (v_z+w_y)h_z + v_y}{\sqrt{1 + h_x^2 + h_z^2}} \\ &= \frac{1}{\Ca} \frac{2h_xh_{xz}h_z - (1+h_z^2)h_{xx} - (1+h_x^2)h_{zz}}{(1 + h_x^2 + h_z^2)^{3/2}},
    \end{split}
    \\
    \label{eqn:3d-stress2}0 &= (v_z+w_y)h_xh_z - (1-h_x^2)(u_y+v_x) + 2(u_x-v_y)h_x + (u_z+w_x)h_z, \\
    \label{eqn:3d-stress3}0 &= (u_y+v_x)h_xh_z + (1-h_z^2)(v_z-w_y) + (u_z-w_x)h_x - 2(v_y-w_z)h_z,
  \end{align}
  where \cref{eqn:3d-stress1} represents the normal stress balance, and \cref{eqn:3d-stress2,eqn:3d-stress3} correspond to the two independent tangential stress balances, respectively. 

  The controls $f(x,z,t)$ act through the boundary conditions at the wall $y=0$:
  \begin{equation}\label{eqn:wallBC}
  u(x, y=0,z,t) = 0, \quad w(x, y=0,z,t) = 0, \quad    v(x, y=0,z,t) = f(x,z,t).
  \end{equation}
  
  Finally, by defining the downstream and cross-stream fluxes $q(x,z,t)$ and $r(x,z,t)$ as follows
\begin{equation}
q(x,z,t) = \int_0^h u(x, y, z, t) \diff y.
    r(x,z,t) = \int_0^h w(x, y,z, t) \diff y,
  \end{equation}
  we obtain the mass conservation equation
  \begin{equation}
    \label{eqn:3d-conservation}h_t + q_x + r_z = f.
  \end{equation}
  
  The equations outlined above are not fundamentally different from their 2D counterparts, but in practice they are much more difficult to solve numerically, which is required for the \textit{in silico} control experiments.

In numerous previous studies, numerical tools based on the finite volume method, be they developments based on in-house code infrastructure or widely used open-source packages such as Gerris~\cite{popinet2003gerris} and Basilisk~\cite{popinet2025basilisk}, have been used successfully in two-dimensional contexts (e.g. \cite{denner2018solitary, cimpeanu2021active, holroyd2024linear}). Extending this workflow to three dimensions, however, can rapidly become computationally restrictive, and arguably does not warrant the added computational burden given that phenomena such as rupture (dewetting or overturning nonlinear waves) or coalescence, which benefit from the interface-capturing machinery offered in the finite volume and attached volume-of-fluid methodology context, are not considered herein. Consequently, we instead opt to implement a dedicated numerical architecture (from formulation to meshing strategy, to the numerical linear algebra setup) for the present study using the oomph-lib~\cite{heil2006oomph} finite-element framework, which  provides greater flexibility for the high-order spatial discretisations required in the present setting. We solve the weak form of the Navier-Stokes equations using the Galerkin formulation. The problem is discretised on a spine mesh, constructed from columns of quad Taylor-Hood elements whose nodes are free to move in the vertical direction \(y\), but are fixed in the directions parallel to the base, \(x\) and \(z\).  In this proposed configuration, we achieve third-order spatial convergence, allowing the three-dimensional problem to be resolved at a substantially reduced computational cost compared to lower-order discretisations. The code used to generate all results from the present investigation can be found at \url{https://github.com/OaHolroyd/oomph-thin-film-control}.

\subsection{Weighted-residual equations in 3D}%
\label{sec:3d_wr_equations}

To design controls that stabilise the flat interface of a falling liquid film, it is advantageous to work with a reduced-dimension approximation of the governing equations. Such models describe either the evolution of the interface alone, or the coupled evolution of the interface and the fluxes.
With these reduced-dimension models, we can perform linear stability analysis to predict regions of the parameter space where a flat interface is unstable, providing a first estimate of the number of actuators required to influence the unstable modes.
This is also crucial to design feedback controls, as we will show in Section~\ref{sec:lqr_control}. We will therefore follow the methodology developed in \cite{holroyd2024linear}, where feedback controls were derived using a linear quadratic regulator (LQR) for the weighted-residual approximation of the interface, and subsequently applying this feedback rule to observations of the interface in a DNS of the system (our \emph{in silico} experiment). We note that we have a whole hierarchy of models available, but given the successful application of weighted-residual-based controls in previous work, together with its relatively low computational cost compared to DNS, we focus here on the weighted-residual model.

The original derivation of the three-dimensional weighted-residual model without boundary forcing was performed by \doublecite{ruyer2000improved}, and in 2D with boundary forcing by \doublecite{thompson2016falling}. Here we adapt the first-order derivation in three dimensions in~\cite[Section 5]{ruyer2000improved}, now including the basal forcing term \(f\).

The first step is to introduce the long-wave scaling \(\epsilon = 1/L \ll 1\) according to
  \begin{equation}
  \label{egn:3d-scaling}%
    \begin{gathered}
      X = \epsilon x, \quad Z = \epsilon z, \quad T = \epsilon t, \quad \Ca = \epsilon^2\hat{\Ca}, \\
      v = \epsilon V, \quad w = \epsilon W, \quad f = \epsilon F,
    \end{gathered}
  \end{equation}
  where $L$ denotes the characteristic long-wave length scale in both horizontal directions. Truncating at \(O(\epsilon^2)\), \cref{eqn:3d-momentum1,eqn:3d-momentum2,eqn:3d-momentum3} become
  \begin{align}
    \label{eqn:3d-mom1-lw}\Rey \epsilon (u_T + uu_X + Vu_y) &= -\epsilon p_X + 2 + u_{yy},\\
    \label{eqn:3d-mom2-lw}0 &= -p_y - 2\cot\theta + \epsilon V_{yy}, \\
    \label{eqn:3d-mom3-lw}0 &= -\epsilon p_Z + \epsilon W_{yy}, \\
    \label{eqn:3d-cont-lw}\epsilon u_X + \epsilon V_y &= 0,
  \end{align}
  the interfacial boundary conditions (\cref{eqn:3d-stress1,eqn:3d-stress2,eqn:3d-stress3}) become
  \begin{align}
    \label{eqn:3d-stress1-lw}p &= -2\epsilon h_Xu_y + 2\epsilon V_y - \frac{h_{XX}+h_{ZZ}}{\hat{\Ca}}, \\
    \label{eqn:3d-stress2-lw}0 &= u_y, \\
    \label{eqn:3d-stress3-lw}0 &= \epsilon W_y,
  \end{align}
  at \(y=h\), and the kinematic condition reads
  \begin{equation}
    \label{eqn:3d-kin-wl}\epsilon h_T + \epsilon q_X + \epsilon r_Z = \epsilon F.
  \end{equation}

  Combining \cref{eqn:3d-stress1-lw} with \cref{eqn:3d-stress2-lw,eqn:3d-cont-lw} we have, at \(y=h\),
  \begin{equation}
    \label{eqn:3d-interim}p = -2\epsilon u_X - \frac{h_{XX}+h_{ZZ}}{\hat{\Ca}}.
  \end{equation}
  Integrating \cref{eqn:3d-mom2-lw} with respect to \(y\) from \(0\) to \(h\) and combining with \cref{eqn:3d-interim} (used as the upper limit of the integral), we obtain the following expression for the pressure
  \begin{equation}
    p = 2(h-y)\cot\theta - \frac{h_{XX}+h_{ZZ}}{\hat{\Ca}}.
  \end{equation}

  Using this expression for the pressure and integrating \cref{eqn:3d-mom3-lw} twice with respect to \(y\) (with no slip and \cref{eqn:3d-stress3-lw} as the lower and upper boundary conditions, respectively) gives an expression for the spanwise velocity \(W\), eliminating it from the problem:
  \begin{equation}
    W = \frac{1}{2}\left( 2\cot\theta - \frac{h_{XXZ}+h_{ZZZ}}{\hat{\Ca}} \right)\left( \bar{y}^2 - 2\bar{y} \right),
  \end{equation}
  where \(\bar{y} = y/h\). This means that the spanwise flux can be written explicitly as
  \begin{equation}
    \label{eqn:spanwise-flux}r = - \frac{1}{3}h^3\left(2\cot\theta h_Z - \frac{h_{XXZ}+h_{ZZZ}}{\hat{\Ca}} \right),
  \end{equation}
meaning the spanwise flux may be determined diagnostically from the interfacial height and does not require a separate evolution equation.

  We can eliminate the vertical velocity \(V\) by integrating the continuity condition and using the basal boundary condition \(V(X,0,Z,t) = F(X,Z,T)\):
  \begin{equation}
    V = F(X,Z,T) - \int_0^y u_X + \epsilon W_Z \diff y.
  \end{equation}

  Following \doublecite{ruyer2000improved}, we expand the streamwise velocity as
  \begin{equation}
    u(X,y, Z, T) = \sum_{j=0}^{\infty} a_j(X,Z,T) \phi_j(\bar{y}),
  \end{equation}
  where the basis functions \(\phi_j(z) = z^{j+1} - \frac{j+1}{j+2}z^{j+2}\) are chosen to fulfil the top and bottom boundary conditions: no slip \(u(x,0, z, t) = 0\) and zero tangential stress \(\partial_y u(x,h, z, t) = 0\).

  Explicitly computing the flux, \(q\), we have
  \begin{equation}
  \begin{split}
    q &= \int_{0}^{h} u \diff y, \\
    &= \int_{0}^{h} \sum_{j=0}^{\infty} a_j \phi_j(\bar{y}) \diff y, \\
    &= \frac{1}{3}ha_0 + \sum_{j=0}^{\infty} \frac{2}{(j+2)(j+3)}a_j.
  \end{split}
  \end{equation}
  Discarding all terms with \(j > 0\) (for which \(a_j = O(\epsilon)\) or smaller~\cite{ruyer2000improved}) we therefore have an expression for the final velocity component
  \begin{equation}
    u = \frac{3q}{h}\phi_0(\bar{y}) + O(\epsilon).
  \end{equation}

  Finally, we integrate \cref{eqn:3d-kin-wl,eqn:3d-mom1-lw} with respect to \(y\) over the height of the film against \(\phi_0\) and, using the expressions for \(p\), \(u\), \(V\), and \(r\) in terms of \(y\), \(h\), and \(q\), we arrive at the 3D weighted-residual equations with basal forcing:
  \begin{align}
    \label{eqn:3d-wr-h}%
    \begin{split}
      h_t &= f - q_x + 2\cot\theta h^2h_z^2 + \frac{2}{3}\cot\theta h^3h_z \\
      &\quad-\frac{1}{3\Ca}h^3h_{xxzz} - \frac{1}{\Ca}h^2h_zh_{xxz} - \frac{1}{\Ca}h^2h_zh_{zzz} - \frac{1}{3\Ca}h^3h_{zzzz},
    \end{split}
    \\
    \label{eqn:3d-wr-q}%
    \begin{split}
      q_t &= \frac{1}{2}\frac{q}{h}f + \frac{9}{7}\frac{q^2h_x}{h^2} - \frac{17}{7}\frac{qq_x}{h}
      + \frac{5}{3\Rey}h - \frac{5}{3\Rey}\cot\theta hh_x - \frac{5}{2\Rey}\frac{q}{h^2} \\
      &\quad+ \frac{5}{6\Ca\Rey}hh_{xxx} + \frac{5}{6\Ca\Rey}hh_{xzz},
    \end{split}
  \end{align}
  where we have returned to the original scaling.

  We note that, at first order, the spanwise flux $r$ does not contribute strongly enough to justify the introduction of a separate evolution equation. 
  Instead, its effects enter indirectly through \cref{eqn:3d-wr-h}, where the explicit expression \eqref{eqn:spanwise-flux} is differentiated and substituted into the mass conservation \eqref{eqn:3d-kin-wl}. Consequently, the three-dimensional weighted-residual system retains the same number of evolution equations as its two-dimensional counterpart. 
  \Cref{eqn:3d-wr-q}, governing the streamwise flux $q$, differs from the two-dimensional model (\cite{thompson2016falling}) primarily through the additional surface-tension contribution arising from variations in the spanwise direction.

\section{Stability analysis and control design}%
\label{sec:stability_analysis}

The first step to control design is linear stability analysis. By linearising the weighted-residuals PDEs \cref{eqn:3d-wr-h,eqn:3d-wr-q}, and introducing a small perturbation, we can predict which wavenumber perturbations are unstable and can therefore grow under the time evolution of our interfacial dynamics. These unstable wavenumbers are precisely the modes that must be stabilised by the controller, and therefore provide important guidance for control design. In this section, we first perform linear stability analysis of our problem, and then describe the control design method that we will employ.

\subsection{Linear stability}\label{sec:linstab}
  We follow a similar procedure as in~\cite{holroyd2024linear} to linearise the system of \cref{eqn:3d-wr-h,eqn:3d-wr-q}: 
  we assume that $h = 1+\delta\tilde{h}$ and $q = 2/3 + \delta\tilde {q}$ for small $\delta$, substitute these expressions in \cref{eqn:3d-wr-h,eqn:3d-wr-q}, retain terms of $O(\delta)$, and drop the tildes to obtain the linearised system:
  \begin{align}
    \label{eqn:3d-lin-wr-h}h_t &= f - q_x + \frac{2}{3\Rey}\cot\theta h_{zz} - \frac{1}{3\Ca}h_{xxzz} - \frac{1}{3\Ca}h_{zzzz}, \\
    \label{eqn:3d-lin-wr-q}q_t &= \frac{1}{3}f + \frac{4}{7}h_x - \frac{34}{21}q_x + \frac{5}{\Rey}h - \frac{5}{3\Rey}\cot\theta h_x - \frac{5}{2\Rey}q + \frac{5}{6\Ca\Rey}h_{xxz} + \frac{5}{6\Ca\Rey}h_{xzz}.
  \end{align}

  The resulting linear system has constant coefficients and is therefore amenable to a standard Fourier-mode stability analysis. Therefore, to understand how we might expect the difficulty of the control problem to change as the physical parameters \(\Rey\), \(\Ca\), \(\theta\), \(L_x\), and \(L_z\) vary, we insert a unimodal perturbation of the form
  \begin{equation}
    h = \hat{h} e^{\lambda t + (kx+lz)i}, \qquad q = \hat{q} e^{\lambda t + (kx+lz)i},
  \end{equation}
  into \cref{eqn:3d-lin-wr-h,eqn:3d-lin-wr-q}. We can then link the growth rate \(\lambda\) and the streamwise and spanwise wavenumbers \(k\) and \(l\) through the dispersion relation
  \begin{equation}
    \label{eqn:lam-k-l}%
    \begin{split}
      % quadratic in \lambda
      0 &= 126 \Ca \Rey \, \lambda^2 \\
      % linear in \lambda
      &\quad+\left(
      42 \Rey \, l^2 (k^2 + l^2)
      + 84 \Ca \Rey \cot\theta \, l^2
      + 315 \Ca
      - 204 i \Ca \Rey \, k
      \right)\lambda \\
      % constant (real)
      &\quad+ 105 (k^4 + l^4)
      + 210 k^2 l^2
      - 72 \Ca \Rey \, k^2
      + 210 \Ca \cot\theta \, (k^2 + l^2) \\
      % constant (imag)
      &\quad+ 68 i \Rey \, ( k l^4 + k^3 l^2)
      + 136 i \Ca \Rey \cot\theta \, k l^2
      + 630 i \Ca \, k.
    \end{split}
  \end{equation}
  We are interested in the region of \((k, l)\)-space for which the perturbation is unstable, i.e.~\(\Re(\lambda) > 0\). To find this region, we compute the zero growth contour where \(\lambda = i\mu\), \(\mu \in \mathbb{R}\), making \(\lambda\) purely imaginary. This separates \cref{eqn:lam-k-l} into a real and imaginary problem in \(k\), \(l\), and \(\mu\). Eliminating \(\mu\) we are left with the implicit equation
  \begin{equation}
    \label{eqn:critial_contour}%
    \begin{split}
      % 12th order
      0 &= 3920 \Rey^2 \, k^2 l^6 (k^4 + l^4)
      + 5880 \Rey^2 \, k^4 l^8
      + 980 \Rey^2 \, l^4 (k^8 + l^8) \\
      % 10th order
      &\quad+ 17640 \Ca \Rey^2 \cot\theta \, k^2 l^6 (k^2 + l^2)
      - 1344 \Ca \Rey^3 \, k^4 l^6 \\
      &\quad+ 5880 \Ca \Rey^2 \cot\theta \, l^4 (k^6 + l^6)
      - 672 \Ca \Rey^3 \, k^2 l^4 (k^4 + l^4) \\
      % 8th order
      &\quad+ 11760 \Ca^2 \Rey^2 \cot^2\theta \, l^4 (k^4 + l^4)
      - 2688 \Ca^2 \Rey^3 \cot\theta \, k^2 l^4 (k^2 + l^2) \\
      &\quad+ 14700 \Ca \Rey \, l^2 (k^6 + l^6)
      + 23520 \Ca^2 \Rey^2 \cot^2\theta \, k^2 l^6
      + 44100 \Ca \Rey \, k^2 l^4 (k^2 + l^2) \\
      % 6th order
      &\quad+ 58800 \Ca^2 \Rey \cot\theta \, k^4 l^2 (k^4 + l^4)
      + 7840 \Ca^3 \Rey^2 \cot^3\theta \, l^4 (k^2 + l^2) \\
      &\quad+ 117600 \Ca^2 \Rey \cot\theta \, k^2 l^4
      - 15520 \Ca^2 \Rey^2 \, k^2 l^2 (k^2 + l^2)
      - 2688 \Ca^3 \Rey^3 \cot^2\theta \, k^2 l^4 \\
      % 4th order
      &\quad+ 58800 \Ca^3 \Rey \cot^2\theta \, l^2 (k^2 + l^2)
      + 110250 \Ca^2 \, k^2 l^2 \\
      &\quad- 31040 \Ca^3 \Rey^2 \cot\theta \, k^2 l^2
      + 55125 \Ca^2 \, (k^4 + l^4) \\
      % 2th order
      &\quad+ 110250 \Ca^3 \cot\theta \, (k^2 + l^2)
      - 88200 \Ca^3 \Rey \, k^2.
    \end{split}
  \end{equation}

  This expression is clearly too complex to draw meaningful information from. However, if we set \(l = 0\), which corresponds to purely streamwise waves, we recover the same stability threshold that was determined in~\cite{holroyd2024linear}.

  \begin{figure}[h!]
    \centering
    \begin{subfigure}{0.7\textwidth}
      \centering
      \begin{tikzpicture}
        \begin{axis}[
          width=\textwidth,
          % height=1.0\textheight,
          unit vector ratio=1 1 1,
          unit rescale keep size=unless limits declared,
          ylabel = {\(l\)},
          xmin = 0,
          xmax = 5,
          ymin = 0,
          ymax = 2,
          xtick distance={1},
          ytick distance={1},
          xticklabels={},
          legend style={at={(1.02,0.5)},anchor=west}
          ]

          \addplot[color=black, only marks, mark=x,forget plot]
          table[x=k, y=l]
          {data/critical-contour/out/kl-grid.dat};

          % Varying Re
          \addplot[color=red, solid]
          table[x=k, y=l]
          {data/critical-contour/out/kl-4.0-0.025-32-32-1.047198.dat};
          \addlegendentry{\(\Rey = 4\)}

          \addplot[color=red, densely dashdotdotted]
          table[x=k, y=l]
          {data/critical-contour/out/kl-8.0-0.025-32-32-1.047198.dat};
          \addlegendentry{\(\Rey = 8\)}

          \addplot[color=red, densely dashdotted]
          table[x=k, y=l]
          {data/critical-contour/out/kl-12.0-0.025-32-32-1.047198.dat};
          \addlegendentry{\(\Rey = 12\)}

          \addplot[color=red, densely dashed]
          table[x=k, y=l]
          {data/critical-contour/out/kl-16.0-0.025-32-32-1.047198.dat};
          \addlegendentry{\(\Rey = 16\)}

          \addplot[color=red, densely dotted]
          table[x=k, y=l]
          {data/critical-contour/out/kl-20.0-0.025-32-32-1.047198.dat};
          \addlegendentry{\(\Rey = 20\)}
        \end{axis}
      \end{tikzpicture}
    \end{subfigure}
    \begin{subfigure}{0.7\textwidth}
      \centering
      \begin{tikzpicture}
        \begin{axis}[
          width=\textwidth,
          % height=1.0\textheight,
          unit vector ratio=1 1 1,
          unit rescale keep size=unless limits declared,
          ylabel = {\(l\)},
          xmin = 0,
          xmax = 5,
          ymin = 0,
          ymax = 2,
          xtick distance={1},
          ytick distance={1},
          xticklabels={},
          legend style={at={(1.02,0.5)},anchor=west}
          ]

          \addplot[color=black, only marks, mark=x, forget plot]
          table[x=k, y=l]
          {data/critical-contour/out/kl-grid.dat};

          % Varying Ca
          \addplot[color=blue, solid]
          table[x=k, y=l]
          {data/critical-contour/out/kl-12.0-0.010-32-32-1.047198.dat};
          \addlegendentry{\(\Ca = 0.01\)}

          \addplot[color=blue, densely dashdotdotted]
          table[x=k, y=l]
          {data/critical-contour/out/kl-12.0-0.020-32-32-1.047198.dat};
          \addlegendentry{\(\Ca = 0.02\)}

          \addplot[color=blue, densely dashdotted]
          table[x=k, y=l]
          {data/critical-contour/out/kl-12.0-0.030-32-32-1.047198.dat};
          \addlegendentry{\(\Ca = 0.03\)}

          \addplot[color=blue, densely dashed]
          table[x=k, y=l]
          {data/critical-contour/out/kl-12.0-0.040-32-32-1.047198.dat};
          \addlegendentry{\(\Ca = 0.04\)}

          \addplot[color=blue, densely dotted]
          table[x=k, y=l]
          {data/critical-contour/out/kl-12.0-0.050-32-32-1.047198.dat};
          \addlegendentry{\(\Ca = 0.05\)}
        \end{axis}
      \end{tikzpicture}
    \end{subfigure}
    \begin{subfigure}{0.7\textwidth}
      \centering
      \begin{tikzpicture}
        \begin{axis}[
          width=\textwidth,
          % height=1.0\textheight,
          unit vector ratio=1 1 1,
          unit rescale keep size=unless limits declared,
          xlabel = {\(k\)},
          ylabel = {\(l\)},
          xmin = 0,
          xmax = 5,
          ymin = 0,
          ymax = 2,
          xtick distance={1},
          ytick distance={1},
          legend style={at={(1.02,0.5)},anchor=west}
          ]

          \addplot[color=black, only marks, mark=x, forget plot]
          table[x=k, y=l]
          {data/critical-contour/out/kl-grid.dat};

          % Varying theta
          \addplot[color=green, solid]
          table[x=k, y=l]
          {data/critical-contour/out/kl-12.0-0.025-32-32-0.157080.dat};
          \addlegendentry{\(\theta = \pi / 20\)}

          \addplot[color=green, densely dashdotdotted]
          table[x=k, y=l]
          {data/critical-contour/out/kl-12.0-0.025-32-32-0.471239.dat};
          \addlegendentry{\(\theta = 3\pi / 20\)}

          \addplot[color=green, densely dashdotted]
          table[x=k, y=l]
          {data/critical-contour/out/kl-12.0-0.025-32-32-0.785398.dat};
          \addlegendentry{\(\theta = 5\pi / 20\)}

          \addplot[color=green, densely dashed]
          table[x=k, y=l]
          {data/critical-contour/out/kl-12.0-0.025-32-32-1.099557.dat};
          \addlegendentry{\(\theta = 7\pi / 20\)}

          \addplot[color=green, densely dotted]
          table[x=k, y=l]
          {data/critical-contour/out/kl-12.0-0.025-32-32-1.413717.dat};
          \addlegendentry{\(\theta = 9\pi / 20\)}
        \end{axis}
      \end{tikzpicture}
    \end{subfigure}
    \caption{Upper-right quadrant of the critical stability contour \(\Re(\lambda) = 0\) as \(\Rey\) (top), \(\Ca\) (middle), and \(\theta\) (bottom) vary. Unless otherwise specified, we fix \(\Rey = 12\), \(\Ca = 0.025\), \(\theta = \pi/3\), and $L_x=L_z=32$. If a wavenumber pair lies inside the stability contour it is unstable. As in the 2D case, increasing any of the three parameters makes the system more unstable.}%
    \label{fig:critical_contour}
  \end{figure}
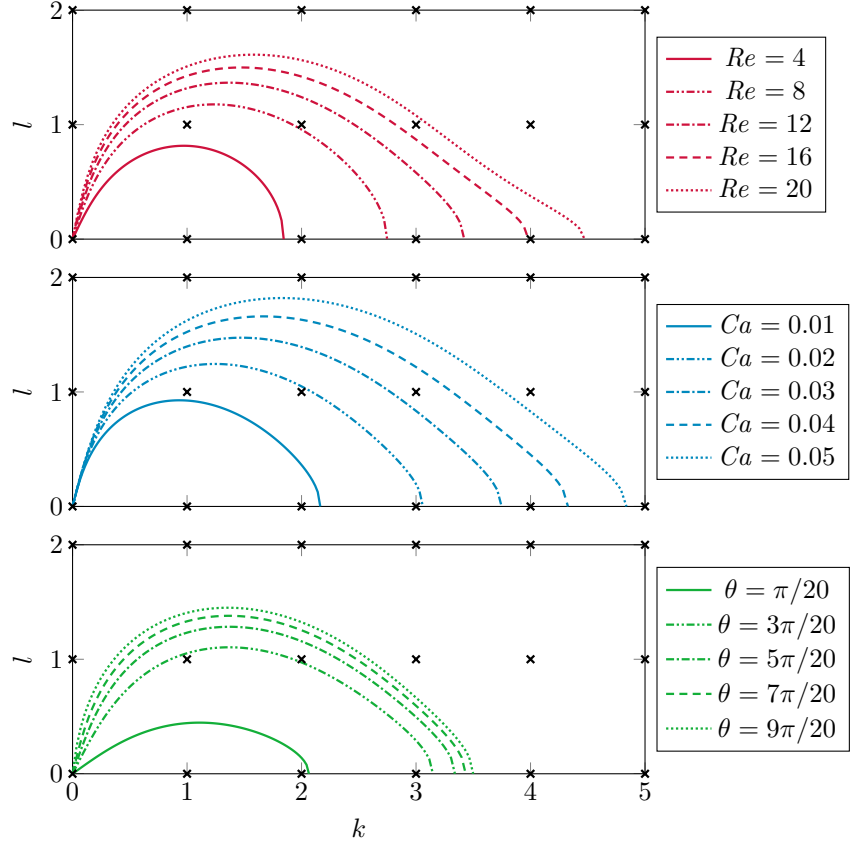

  To better understand the effects of changing \(\Rey\), \(\Ca\), and \(\theta\) on the stability, we turn to \cref{fig:critical_contour}, where we show how the stability threshold described by \cref{eqn:critial_contour} changes as the three parameters increase. For sufficiently small values of any of the parameters, we only see instabilities appearing in streamwise waves, but when the combination of the parameters is sufficiently large, we see the inclusion of wavemodes with a spanwise component in the unstable region. While these more complex waves are now unstable, and therefore could emerge in simulations of the uncontrolled film, in practice we only see the purely streamwise \((1,0)\)-wave appearing, since it has by far the largest growth rate associated with it: the spanwise interfacial oscillations are, like the streamwise modes, opposed by surface tension, but they do not have the same gravitational forcing to drive their growth. We will see this later in \cref{fig:3d_interfaces_re,fig:3d_interfaces_m} at \(t=0\): although the initial condition is set to a small \((1,1)\)-perturbation, after \(100\) non-dimensional time units, the \((1,0)\)-wave dominates. The number of these new unstable modes, which are on a 2D lattice rather than on a line, can grow much faster than in the 2D problem, and so we would expect to find that many more actuators are required than in previous work.

  Note that changing the size of the domain along either the \(x\)- or \(z\)-axis simply scales the unstable region in the same direction (or equivalently expands or compresses the wavenumber pairs in or out of the unstable region).

\subsection{LQR control design}%
\label{sec:lqr_control}

  The derivation and computation of the gain matrix $K$ for the Linear Quadratic Regulator (LQR) control strategy are a straightforward extension of the two-dimensional framework developed in \cite{holroyd2024linear}, and therefore only a brief summary is given here. We formulate the control problem as a Linear Quadratic Regulator (LQR) problem based on the linearisation of the weighted-residual system about the Nusselt solution given in \cref{eqn:3d-lin-wr-h,eqn:3d-lin-wr-q}. Following spatial discretisation, the resulting system can be written in the form
\begin{equation}\label{eqn:lin-disc}
\dot{\mathbf{\xi}}=A\mathbf{\xi}+B\mathbf{\eta}\qquad \mathbf{\xi}(0) = \mathbf{\xi_0},
\end{equation}
where $\mathbf{\xi}$ contains the discretised perturbations to the interfacial height and streamwise flux, $\mathbf{\xi_0}$ discretises the initial conditions, and $\mathbf{\eta}$ denotes the actuator amplitudes. In addition, $A$ contains the discretisation of the linearised operators, while $B$ contains the information on the actuator locations (i.e. it is the discretised version of the functions $d(x,z,t)$ in \cref{eqn:actuator-3d}).

The control objective is to minimise the quadratic cost functional
\begin{equation}\label{eqn:cost}
    \kappa = \int_0^\infty \langle \mathbf{\xi}-\mathbf{\bar{\xi}},Q(\mathbf{\xi}-\mathbf{\bar{\xi}})\rangle + \langle \mathbf{\eta},R\mathbf{\eta}\rangle \diff t,
\end{equation}
where $\bar{\xi}$ denotes the discretised Nusselt state and $Q$ and $R$ are symmetric semi-definite and symmetric positive definite matrices of appropriate sizes, with relative weights providing a balance between the cost of deviating from the Nusselt solution and the cost of controls, respectively. Under the assumption that the control is given by $\mathbf{\eta} = K(\mathbf{\xi}-\mathbf{\bar{\xi}})$, where now we are interested in finding the \emph{gain} matrix $K$, standard LQR theory yields
that $K = -R^{-1}B^T P$, where $P$ solves the continuous algebraic Riccati equation (CARE)
\begin{equation}\label{eqn:Riccati}
A^TP +PA - PB R^{-1}B^TP + Q=0.   
\end{equation}
For more details in this derivation, see \cite{holroyd2024linear}.

By solving \cref{eqn:Riccati} numerically for $P$, we can compute $K$ and therefore have a feedback law to stabilise the flat solution to the linearised problem (and indeed to the full problem, as will be shown in \cref{sec:3d-results}). As in the two-dimensional case, \cref{eqn:Riccati} is a nonlinear equation with $O(n^2)$ unknowns, where $n$ is the size of the discretisation vector $\xi$. Since $n$
must be large enough to provide a good approximation to the continuous problem, numerically solving the Ricatti equation
is non-trivial. Following \cite{holroyd2024linear}, we use the classical eigenvector approach described by \doublecite{macfarlane1963eigenvector}, \doublecite{potter1966matrix}, and \doublecite{vaughan1970nonrecursive}: letting $P = P_2P_1^{-1}$,with square
matrices $P_1$ and $P_2$, and defining the Hamiltonian matrix
\[
H = \left[\begin{array}{cc}A & -B R^{-1}B \\ -Q & -A^T\end{array}\right],
\]
we can rewrite \cref{eqn:Riccati} as
\begin{equation}\label{eqn:e-value}
H \left[\begin{array}{c}P_1 \\ P_2\end{array}\right] = \Lambda \left[\begin{array}{c}P_1 \\ P_2\end{array}\right],
\end{equation}
where $\Lambda$ is the diagonal matrix containing the eigenvalues of $H$. We have therefore reformulated the CARE as an eigenvalue problem, and consequently standard methods can be used to compute
the eigenvalues and eigenvectors of H.

The key distinction with our three-dimensional problem lies its computational complexity. As in direct numerical simulations of the Navier-Stokes equations, the introduction of a spanwise coordinate leads to a substantial increase in the number of degrees of freedom. Consequently, the linearised and discretised versions of the weighted-residual system in \cref{eqn:3d-wr-h,eqn:3d-wr-q} are considerably larger than their two-dimensional counterparts, leading to increased computational costs associated with matrix assembly, storage, and the solution of the large-scale linear algebra problems required to compute the feedback operator.

\begin{figure}[t]
    \centering
    \begin{subfigure}{\textwidth}
      \centering
      \begin{tikzpicture}
        \begin{axis}[
          width=0.8\textwidth,
          % height=1.0\textheight,
          unit vector ratio=1 1 1,
          xlabel = {},
          ylabel = {\(x\)},
          xmin = 0,
          xmax = 64,
          ymin = 0,
          ymax = 32,
          xticklabels={}
          ]

          \addplot graphics [xmin = 0, xmax = 64, ymin = 0, ymax = 32] {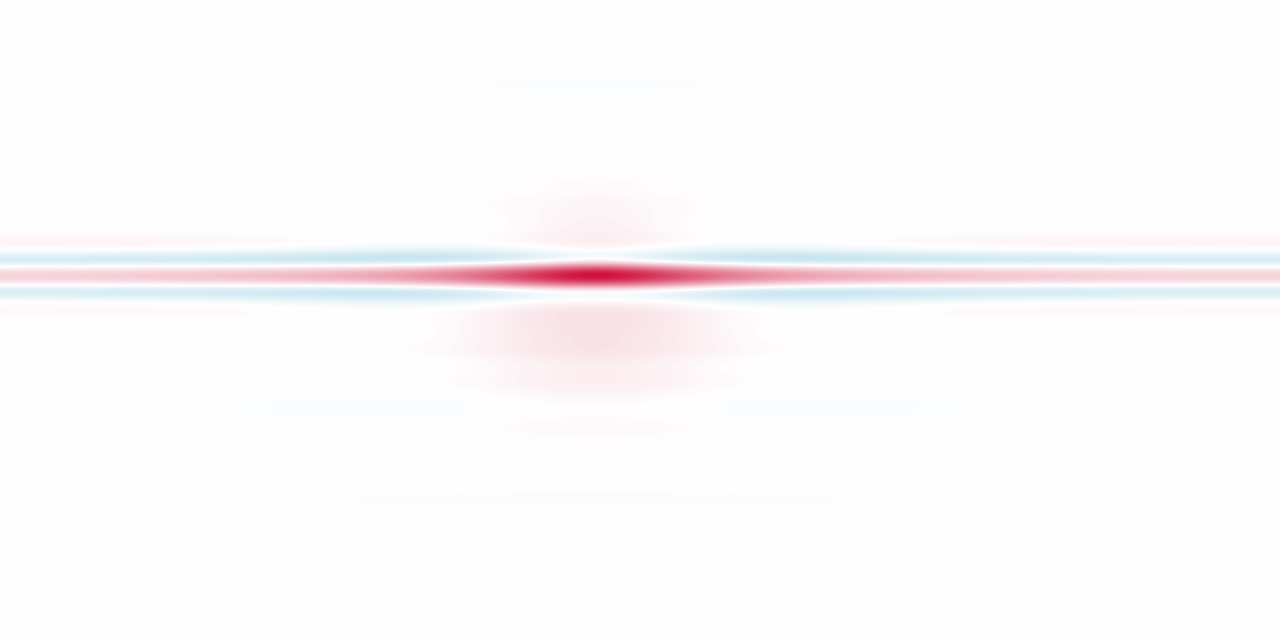};%
        \end{axis}
      \end{tikzpicture}
    \end{subfigure}
    \begin{subfigure}{\textwidth}
      \centering
      \begin{tikzpicture}
        \begin{axis}[
          width=0.8\textwidth,
          % height=1.0\textheight,
          unit vector ratio=1 1 1,
          xlabel = {\(z\)},
          ylabel = {\(x\)},
          xmin = 0,
          xmax = 64,
          ymin = 0,
          ymax = 32,
          ]

          \addplot graphics [xmin = 0, xmax = 64, ymin = 0, ymax = 32] {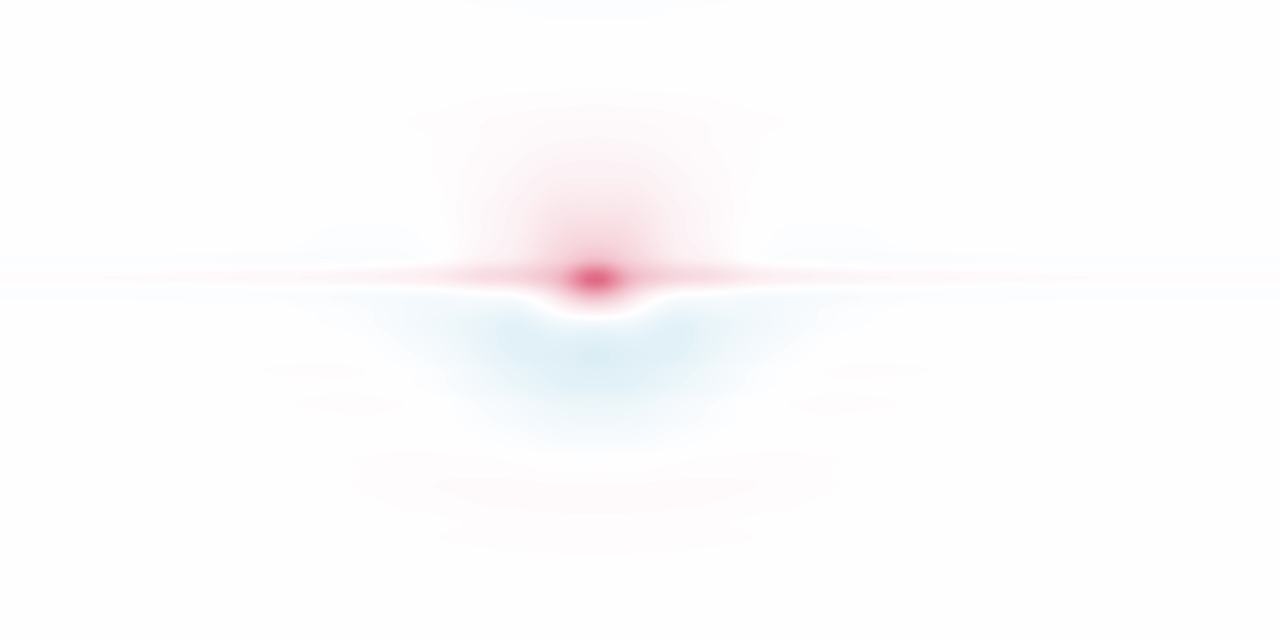};%
        \end{axis}
      \end{tikzpicture}
    \end{subfigure}
    \caption{The gain function \(k(x,z)\) for an actuator located at \((13.75, 29.75)\), with \(\Rey=15\), \(\Ca=0.025\), and \(m=128\) actuators in a regular grid covering the domain. The portion of the gain associated with the interface is shown at the top, and the portion associated with the flux at the bottom. Both are normalised symmetrically around zero using the same factor.}%
    \label{fig:3d-gain}
  \end{figure}

The slowest step in this calculation, taking the vast majority of the total computational time, is the solution of the eigenvalue problem \eqref{eqn:e-value} (here performed using LAPACK's \texttt{zgeev}~\cite{anderson1999lapack}, which is a single-threaded algorithm): since the two-dimensional discretisation contains $O(n^2)$ degrees of freedom when $n$ points are used in each spatial direction, the dense eigensolve scales as $O(n^6)$ in computational cost and $O(n^4)$ in memory. While this is still a one-off computation, the computational effort involves 20 hours and 100GB of RAM for a medium-sized grid (for example the \(64\times128\) grid used in \cref{fig:3d-gain}) rather than seconds to minutes, which was the case in the two-dimensional counterpart. While the time-cost can be coped with simply by waiting longer, memory is a hard limit imposed by the available hardware.

  The increased computational time could be reduced by either computing the eigenvalues in a more efficient manner: either in parallel (using SCALPACK~\cite{blackford1997scalapack} for instance) or by only computing the most stable half of the required eigenvalues (for instance with iterative methods as in~\cite{stewart2002krylov}). A more significant departure from this would be to use approximation methods designed for solving large Ricatti equations, for instance those described by \doublecite{massei2024data} for banded problems (such as those that arise as a result of discretising PDEs).
  
\begin{figure}[htb]
    \centering%
    \begin{tikzpicture}
      \begin{axis}[
        width=0.95\textwidth,
        height=0.6\textwidth,
        axis lines = box,
        xlabel = {\(x\)},
        ylabel = {Feedback gain (scaled)},
        ymin = -1,
        ymax = 2,
        xmin = 0,
        xmax = 32,
        xtick={0, 5, 10, 15, 20, 25, 30},
        legend cell align=left
        ]

        \addplot [
        color = black, solid, smooth
        ]
        table[x expr=\thisrowno{0}, y expr=4*\thisrowno{2}]
        {data/K-3d/k-64x128-slice.dat};
        \addlegendentry{3D gain slice (64 points)}

        \addplot [
        color = red, densely dashed, smooth
        ]
        table[x expr=\thisrowno{0}, y expr=4*\thisrowno{1}+0.25]
        {data/K-3d/k-2d.dat};
        \addlegendentry{2D gain (64 points)}

        \addplot [
        color = blue, densely dotted, smooth
        ]
        table[x expr=\thisrowno{0}, y expr=12*\thisrowno{1}]
        {data/K-3d/k-2d-HR.dat};
        \addlegendentry{2D gain (400 points)}
      \end{axis}
    \end{tikzpicture}
    \caption{Streamwise slice through the three-dimensional interface-height gain in \cref{fig:3d-gain}, compared with two-dimensional gains at matching and higher resolution. The three-dimensional gain retains the same local structure, although its trough–peak–trough feature is less pronounced.
    Only the portions of the gain associated with the interface are shown.}%

    \label{fig:3d_gain_slice}
  \end{figure}
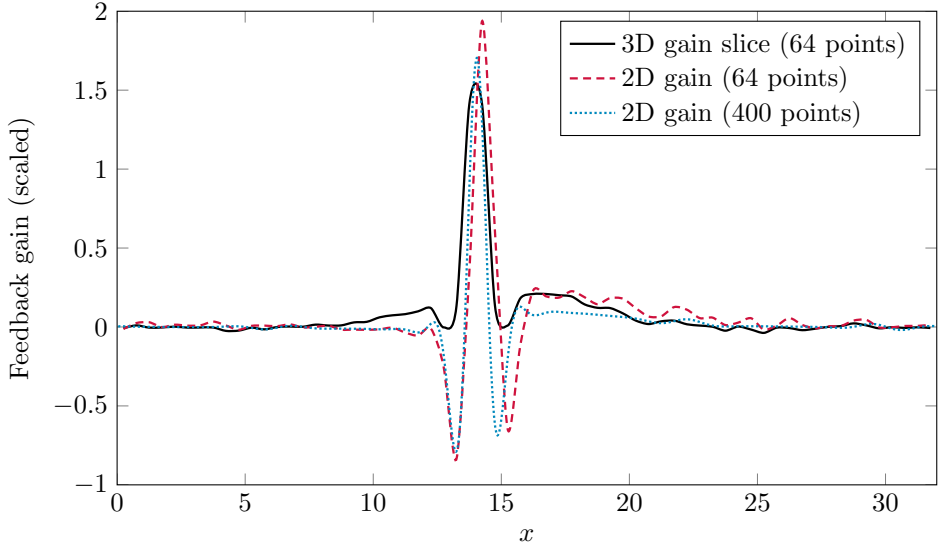

  Each row of the gain matrix \(K\) that results from solving the Ricatti equation is a discrete representation of a function \(k(x,z)\) denoting the
  weight assigned to an interfacial measurement at a location \((x, z)\) to one of the actuators:
  \begin{equation}
    a_i(t) = \int_{0}^{L_x} \int_{0}^{L_z} k(x,z) \cdot \xi(x,z,t) \diff z \diff x \approx K_i \xi.
  \end{equation}
  In \cref{fig:3d-gain} we can see that, like in the 2D case, the regions nearest the actuator are the most significant in determining the actuator strength. Although, due to the cost of computing it,
  the 3D gain matrix has a much lower resolution than its 2D counterpart (computed previously in \cite{holroyd2024linear}), \cref{fig:3d_gain_slice} shows that there are similarities in the structure: for example, the trough-peak-trough feature that arises from the second derivative appearing in the matrix \(P\) which solves the Ricatti equation. This pattern extends along the entire width of the domain, suggesting that the purely-streamwise gravity-driven  \((k,0)\)-wavemodes are the most significant, as we would expect from the scaling in \cref{egn:3d-scaling}. The flux component of the gain is weaker and, although we still see the same shape as the 2D case along the streamwise axis through the actuator position, it lacks the cross-stream symmetry of the interfacial component. 
  One possible explanation is that surface-tension effects make interfacial contributions from across the full span of the domain more significant to the dynamics, whereas the flux contributions remain comparatively localised.
  
Once the gain matrix $K$ has been computed, the control amplitudes are obtained by applying the feedback law to observations of the interface (and, where available, the flux). In Section 5 these controls are applied directly to observations from DNS of the Navier-Stokes equations, demonstrating the robustness of the resulting feedback law across the model hierarchy.

\section{Numerical results}%
\label{sec:3d-results}
Motivated by the success of the corresponding two-dimensional framework, we now test the control methodology directly in DNS of the three-dimensional Navier-Stokes equations.

Throughout this section, we consider parameter values relevant to practical applications, with Reynolds numbers in the range $12 \leq \Rey \leq 80$, $\Ca=0.025$, and $\theta=\pi/3$. Unless otherwise stated, we use $L_x=L_z=32$. Simulations are allowed to evolve without control over the interval $t\in[-100,0]$, permitting the film to reach its saturated state before control is activated. At $t=0$, feedback control os activated according to  $a_i(t) = K\left[(h^*(x,z,t)-1); (q^*(x,t)-2/3\right]$,  where $h^*(x,z,t)$ are observations of the interface in the corresponding DNS simulation, and the flux $q(x,t)$ is approximated by $q^*(x,t) = \frac{2}{3}h(x,t)$ as previously shown to be effective (\cite{thompson2016stabilising,holroyd2024linear}); furthermore, $K$ is computed for the weighted residuals model as described in \cref{sec:lqr_control}. 

Since spanwise instabilities are possible in the present three-dimensional setting, we initialise the system with a $(1,1)$-mode perturbation, thereby introducing energy into disturbances with both streamwise and spanwise structure. 

In the following subsections, we investigate the effects
of varying the Reynolds number, domain size, actuator number, and actuator arrangement. 
In each case, only one parameter is varied while all others are held fixed, enabling the influence of each factor on the stabilisability of the system to be assessed independently. 
While in some of these cases (e.g. varying the Reynolds number), we expect to have broadly similar results as observed in the two-dimensional problem in \cite{holroyd2024stabilisation}, others, most notably the placement of the actuators, introduce genuinely new challenges that arise only in the three-dimensional setting.

\subsection{Varying Reynolds number}%
\label{sub:varying_reynolds_number}
  The challenges associated with increasing \(\Rey\) are similar to those observed in the 2D setting: we observe an increase in the number of unstable modes, but in the present three-dimensional system, higher Reynolds numbers also permit spanwise instabilities in addition to purely streamwise disturbances (as seen in  \cref{fig:critical_contour}).

  \begin{figure}[htbp]
    \centering%
    \begin{tikzpicture}
      \begin{groupplot}[
        group style={
            group name=my plots,
            group size=4 by 6,
            xlabels at=edge bottom,
            xticklabels at=edge bottom,
            ylabels at=edge left,
            yticklabels at=edge left,
            vertical sep=6pt,
            horizontal sep=6pt
        },
        width=0.38\textwidth,
        unit vector ratio=1 1 1,
        unit rescale keep size=unless limits declared,
        axis lines = box,
        xlabel = {\(x\)},
        ylabel = {\(z\)},
        ymin = 0,
        ymax = 32,
        xmin = 0,
        xmax = 32,
        clip=false,
        ]

        \providecommand\makeplot[3]{%
          \addplot graphics [xmin = 0, xmax = 32, ymin = 0, ymax = 32] {data/3d-data/#1/images/im-#2.png};%
          \ifnum#2>500%
          \addplot[thin,color=black,only marks,mark=x] table[x=x,y expr=\thisrowno{2} > 0 ? \thisrowno{1} : NaN] {data/3d-data/#1/actuators/ac-#2.dat};%
          \addplot[thin,color=black,only marks,mark=o] table[x=x,y expr=\thisrowno{2} < 0 ? \thisrowno{1} : NaN] {data/3d-data/#1/actuators/ac-#2.dat};%
          \else
          \node[above] at (axis cs: 16,32) {#3};%
          \fi%
        }

        \providecommand\makerow[1]{%
          \nextgroupplot%
          \makeplot{output-12-64x64}{#1}{$\Rey = 12$}%
          \nextgroupplot%
          \makeplot{output-20-64x64}{#1}{$\Rey = 20$}%
          \nextgroupplot%
          \makeplot{output-40-64x64}{#1}{$\Rey = 40$}%
          \nextgroupplot%
          \makeplot{output-80-64x64}{#1}{$\Rey = 80$}%
          \pgfmathsetmacro{\time}{round((0.2*#1-100))} \node[right] at (axis cs: 32,16) {$t=\pgfmathprintnumber{\time}$};
          %\FPeval{\time}{clip(0.2*#1 - 100)}%
          %\node[right] at (axis cs: 32,16) {$t = \time$};%
        }

        \makerow{500}
        \makerow{510}
        \makerow{550}
        \makerow{650}
        \makerow{850}
        \makerow{1000}
      \end{groupplot}
    \end{tikzpicture}
    \caption{Series of 3D DNS interfacial snapshots of the film for different values of \(\Rey\) as it evolves after the activation of the controls at \(t=0\). Symbols represent the actuator locations: crosses represent removal of fluid (negative vertical velocity/moving into the page), and circles represent injection of fluid (positive vertical velocity/moving out of the page). Other parameters are kept fixed: \(\Ca = 0.025\), \(\theta = \pi / 3\).}%
    \label{fig:3d_interfaces_re}
  \end{figure}
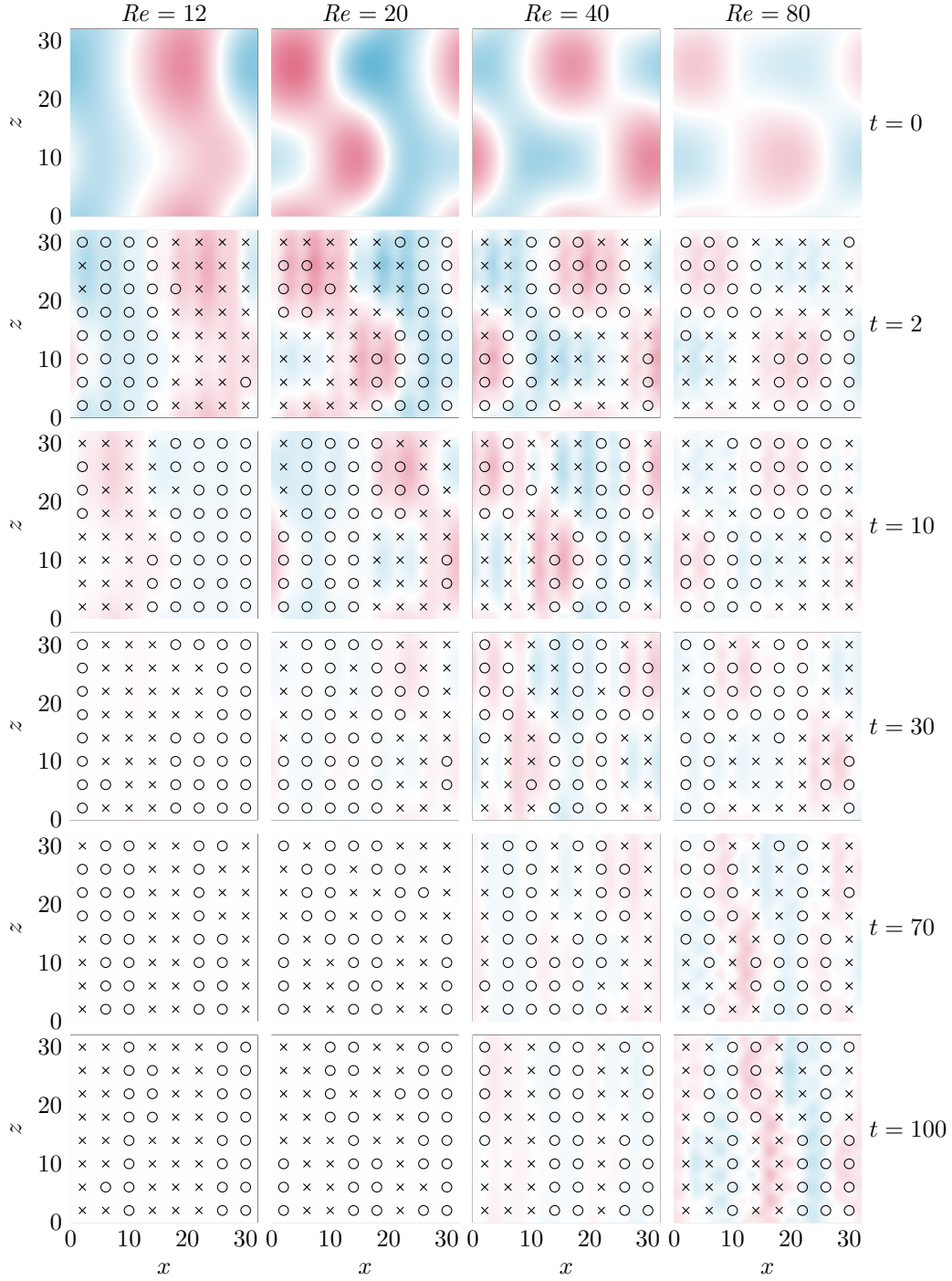

  In \cref{fig:3d_interfaces_re}, we investigate the effect of feedback control by providing snapshots of the resulting film evolution at a selection of time steps. We observe that an evenly spaced $8\times8$ actuator array successfully stabilises the flat film for Reynolds numbers of \(12\), \(20\), and \(40\). As \(\Rey\) increases, the damping rate reduces, until eventually control effectiveness decreases and is lost between \(\Rey=40\) and \(\Rey = 60\). In the final column of \cref{fig:3d_interfaces_re} we can see how, at \(\Rey = 80\), although there is an initial reduction in the size of the perturbation after the activation of the controls, by \(t=100\) an uncontrolled \((2,0)\)-wave has developed.

  \begin{figure}[htb]
    \centering
    \begin{tikzpicture}
      \begin{axis}[
        width=0.95\textwidth,
        % height=1.0\textheight,
        axis lines = box,
        xlabel = {\(t\)},
        ylabel = {\(\norm{h-1}\)},
        ymin = 0.00001,
        xmin = -100,
        xmax = 100,
        ymode = log,
        legend pos=south west,
        ]

        \providecommand\makeplot[2]{%
          \addplot [#2]%
            table[x expr=\thisrowno{0}-100, y expr=\thisrowno{1}]%
            {data/3d-data/#1/results.dat};%
        }

        \makeplot{output-12-64x64}{color=blue,solid,}
        \makeplot{output-20-64x64}{color=blue,densely dashed,}
        \makeplot{output-40-64x64}{color=green,solid,}
        \makeplot{output-60-64x64}{color=green,densely dashed,}
        \makeplot{output-80-64x64}{color=red,solid,}
        \makeplot{output-100-64x64}{color=red,densely dashed,}

        \legend{\(\Rey = 12\), \(\Rey = 20\), \(\Rey = 40\), \(\Rey = 60\), \(\Rey = 80\), \(\Rey = 100\),}
      \end{axis}
    \end{tikzpicture}
    \caption{Norm of the interfacial perturbations over time corresponding to the same simulation setup as \cref{fig:3d_interfaces_re}. After the controls are activated at \(t=0\) we can see that above \(\Rey = 40\) the grid of \(8\times8\) actuators are unable to stabilise the uniform film.}%
    \label{fig:3d_lines_re}
  \end{figure}
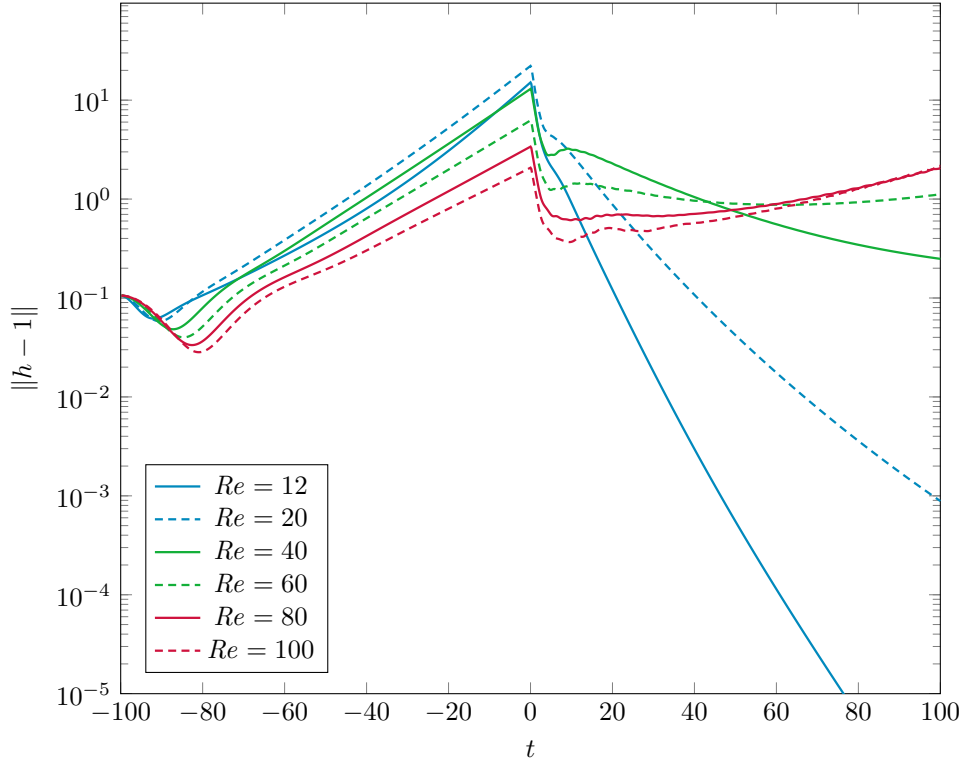

To better see the effect of increasing \(\Rey\), we provide the evolution of $\|h-1\|$ as a function of time for all cases in \cref{fig:3d_lines_re}. We observe that from \(12\) to \(20\) and finally \(40\), stabilisation is achieved, with a reduction in the damping rate clearly apparent. For \(\Rey = 60\), \(80\), and \(100\), complete stabilisation is not achieved, although the growth rates of the unstable modes are significantly reduced. 
One possible explanation for this transition from successful to unsuccessful stabilisation is the change in the number of unstable purely streamwise modes, which increases from seven to eight at \(\Rey = 48\). Above this threshold, the two-dimensional analysis of \cite{holroyd2024linear} suggests that an actuator configuration with only eight streamwise locations approaches the limit of what can be stabilised. This observation raises the possibility that, at least for perturbations that are purely streamwise, the stabilisability of the three-dimensional system depends primarily on the number of distinct actuator locations in the $x$-direction rather than on the total number of actuators.

\subsection{Varying the number of actuators}%
\label{sub:varying_number_of_actuators}
  While in the 2D case we could vary only the number of actuators \(m\) along the one-dimensional base, in 3D we can vary the number of actuators independently in the streamwise and spanwise directions, denoted by \(m_x\) and \(m_z\), respectively.  
  We can now take the 3D problem and, rather than increasing the Reynolds number until stabilisation fails, reduce the number of actuators in either direction until stabilisation of the flat interface is no longer achieved.

For this numerical test, we chose the parameter values \(\Rey = 12\), \(\Ca = 0.025\), \(\theta = \pi/3\), \(L_x = 32\), so that the corresponding two-dimensional problem can be stabilised using only two actuators, and we can explore the three-dimensional effects on control success. We now present a series of computational experiments choosing \(m_x\) and \(m_z\) independently, taking values in \({2, 4, 8}\) (so that the case with the largest number of actuators, \((m_x, m_z) = (8, 8)\), matches the corresponding set of parameters in the previous section).

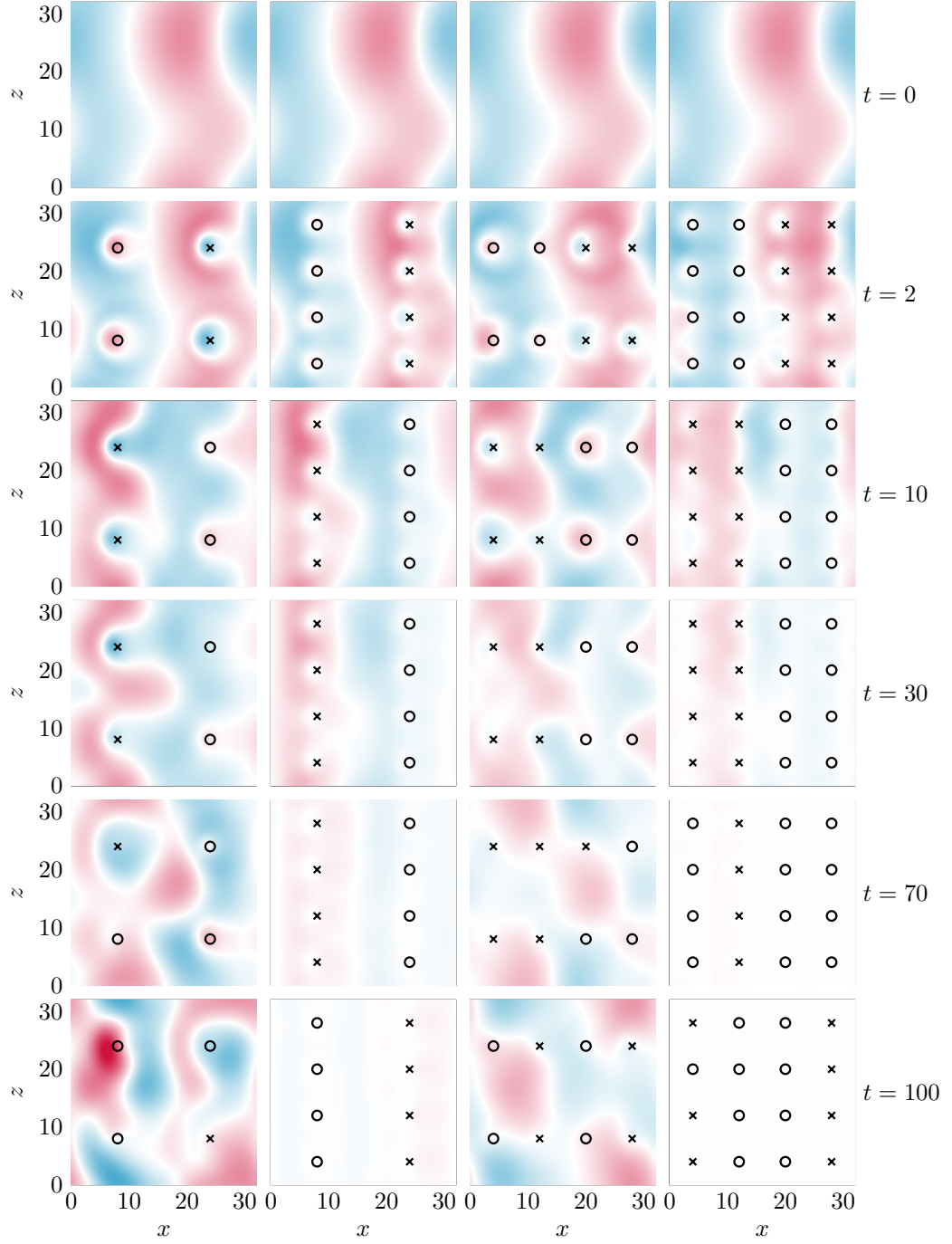
\begin{figure}[htbp]
    \centering%
    \begin{tikzpicture}
      \centering%
      \begin{groupplot}[
        group style={
            group name=my plots,
            group size=4 by 6,
            xlabels at=edge bottom,
            xticklabels at=edge bottom,
            ylabels at=edge left,
            yticklabels at=edge left,
            vertical sep=6pt,
            horizontal sep=6pt
        },
        width=0.38\textwidth,
        unit vector ratio=1 1 1,
        unit rescale keep size=unless limits declared,
        axis lines = box,
        xlabel = {\(x\)},
        ylabel = {\(z\)},
        ymin = 0,
        ymax = 32,
        xmin = 0,
        xmax = 32,
        clip=false,
        ]

        \providecommand\makeplot[2]{%
          \addplot graphics [xmin = 0, xmax = 32, ymin = 0, ymax = 32] {data/3d-data/#1/images/im-#2.png};%
          \ifnum#2>500%
          \addplot[color = black, only marks, mark=x] table[x=x, y expr=\thisrowno{2} > 0 ? \thisrowno{1} : NaN] {data/3d-data/#1/actuators/ac-#2.dat};%
          \addplot[color = black, only marks, mark=o] table[x=x, y expr=\thisrowno{2} < 0 ? \thisrowno{1} : NaN] {data/3d-data/#1/actuators/ac-#2.dat};%
          \fi%
        }

        \providecommand\makerow[1]{%
          \nextgroupplot%
          \makeplot{output-2x2}{#1}%
          \nextgroupplot%
          \makeplot{output-2x4}{#1}%
          \nextgroupplot%
          \makeplot{output-4x2}{#1}%
          \nextgroupplot%
          \makeplot{output-4x4}{#1}%
        \pgfmathsetmacro{\mytime}{round((0.2*#1-100))} \node[right] at (axis cs: 32,16) {$t=\pgfmathprintnumber{\mytime}$};
          
         % \FPeval{\mytime}{clip(0.2*#1 - 100)}%
         % \node[right] at (axis cs: 32,16) {$t = \mytime$};%
        }

        \makerow{500}
        \makerow{510}
        \makerow{550}
        \makerow{650}
        \makerow{850}
        \makerow{1000}
      \end{groupplot}
    \end{tikzpicture}
    \caption{Keeping parameters fixed at \(\Rey=12\), \(\Ca=0.025\), \(\theta=\pi/3\) (corresponding to the red dot-dashed line in \cref{fig:critical_contour}), 3D DNS interfacial snapshots show the effects of different numbers and arrangements of actuators on the stabilisability of the film.}%
    \label{fig:3d_interfaces_m}
  \end{figure}

In \cref{fig:3d_interfaces_m} we present four representative examples of the time evolution of the film, which show us that the effects of varying both the number and arrangement of the actuators are significant. Notably, neither the \(2\times2\) nor the \(4\times2\) arrays of actuators are able to stabilise the 3D problem, despite \(2\) actuators being sufficient for the corresponding 2D version. In the \(4\times2\) case, we can see how the activation of the controls has broken up the \((1, 0)\)-wave, but the spanwise spacing between the two rows is large enough for an uncontrolled \((1, 1)\)-wave to fit between the gaps. For these parameters, we require additional actuators in the spanwise direction to stabilise the film (see the \(2\times4\) and the \(4\times4\) cases). 
This observation is consistent with the stability contour shown in \cref{fig:critical_contour}. The red dot-dashed curve encloses both the $(1,1)$ and $(1,2)$ modes, which, together with their six reflected counterparts in the other three quadrants (increasing the number of unstable modes from 7 to 15), can only be accounted for using more actuators along the \(z\)-axis.

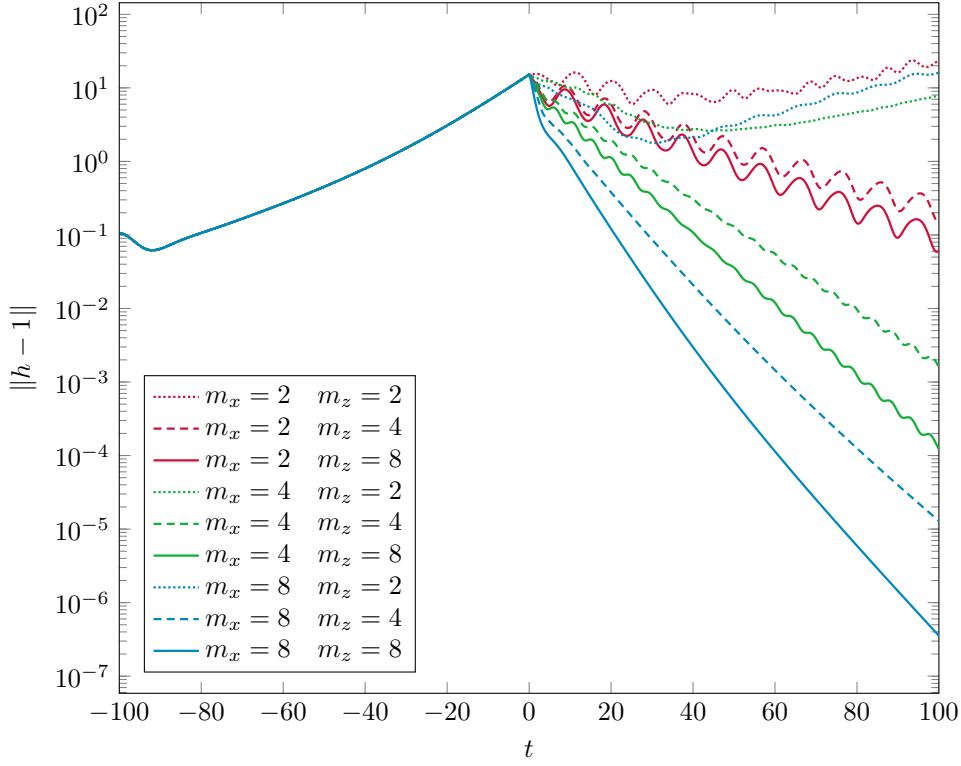
\begin{figure}[htb]
    \centering
    \begin{tikzpicture}
      \begin{axis}[
        width=0.95\textwidth,
        % height=1.0\textheight,
        axis lines = box,
        xlabel = {\(t\)},
        ylabel = {\(\norm{h-1}\)},
        xmin = -100,
        xmax = 100,
        ymode = log,
        legend pos=south west,
        ]

        \providecommand\makeplot[2]{%
          \addplot [#2]%
            table[x expr=\thisrowno{0}-100, y expr=\thisrowno{1}]%
            {data/3d-data/#1/results.dat};%
        }

        \makeplot{output-2x2}{color=red,densely dotted,}
        \makeplot{output-2x4}{color=red,densely dashed,}
        \makeplot{output-2x8}{color=red,solid,}

        \makeplot{output-4x2}{color=green,densely dotted,}
        \makeplot{output-4x4}{color=green,densely dashed,}
        \makeplot{output-4x8}{color=green,solid,}

        \makeplot{output-8x2}{color=blue,densely dotted,}
        \makeplot{output-8x4}{color=blue,densely dashed,}
        \makeplot{output-8x8}{color=blue,solid,}

        \legend{
          \(m_x=2\quad m_z=2\),
          \(m_x=2\quad m_z=4\),
          \(m_x=2\quad m_z=8\),
          \(m_x=4\quad m_z=2\),
          \(m_x=4\quad m_z=4\),
          \(m_x=4\quad m_z=8\),
          \(m_x=8\quad m_z=2\),
          \(m_x=8\quad m_z=4\),
          \(m_x=8\quad m_z=8\)
        }
      \end{axis}
    \end{tikzpicture}
    \caption{Norm of the interfacial perturbations over time for the same parameters as \cref{fig:3d_interfaces_m}. Actuators are switched on at \(t=0\). Stabilisation is only achieved in cases with \(m_z > 2\), with larger \(m_z\) resulting in faster stabilisation. Increasing \(m_x\) increases both the decay rate and smoothness of the convergence.}%
    \label{fig:3d_lines_m}
  \end{figure}

  We can see the comparison between additional simulations in \cref{fig:3d_lines_m} for the deviation from the flat interface in all cases, and the results match those from \cref{fig:3d_interfaces_m}: all of the attempts with \(m_z=2\) (the dotted lines) fail. In the successful cases, more actuators result in faster, more uniform decay since instabilities are unable to grow as much between successive columns of actuators spaced along the \(x\)-axis. We can see how, when \(m_x\) is kept constant, the addition of more actuators along the \(z\)-axis does result in a faster decay rate but does not decrease the magnitude or period of the oscillations around the overall damping trend.

In the two-dimensional problem, the periodic geometry makes it natural to consider equally spaced actuators. 
In three dimensions, however,
the additional spatial direction introduces new geometrical considerations, and actuator placement becomes a significantly richer problem.
In \cref{fig:3d_interfaces_m_shift} and \cref{fig:3d_lines_m_shift}, we consider a case which was unsuccessful in \cref{fig:3d_lines_m} (namely, $4\times 2$ actuators) and introduce an offset to the actuators so that the resulting grid is rotated at an angle of \(\pi/4\) from the gravity vector, which allows the effects of the actuators to be felt in both directions rather than isolated to rows in the \(x\)-direction and columns in the \(z\)-direction.  
We observe that this offset can be sufficient to transform an unsuccessful control configuration into a successful one, producing performance comparable to that obtained using the $2\times4$ actuator arrangement. The results therefore suggest that, in three dimensions, actuator arrangement may be as important as actuator count.

  \begin{figure}[htbp]
    \centering%
    \begin{tikzpicture}
      \begin{groupplot}[
        group style={
            group name=my plots,
            group size=4 by 6,
            xlabels at=edge bottom,
            xticklabels at=edge bottom,
            ylabels at=edge left,
            yticklabels at=edge left,
            vertical sep=6pt,
            horizontal sep=6pt
        },
        width=0.38\textwidth,
        unit vector ratio=1 1 1,
        unit rescale keep size=unless limits declared,
        axis lines = box,
        xlabel = {\(x\)},
        ylabel = {\(z\)},
        ymin = 0,
        ymax = 32,
        xmin = 0,
        xmax = 32,
        clip=false,
        ]

        \providecommand\makeplot[3]{%
          \addplot graphics [xmin = 0, xmax = 32, ymin = 0, ymax = 32] {data/3d-data/#1/images/im-#2.png};%
          \ifnum#2>500%
          \addplot[thin,color=black,only marks,mark=x] table[x=x,y expr=\thisrowno{2} > 0 ? \thisrowno{1} : NaN] {data/3d-data/#1/actuators/ac-#2.dat};%
          \addplot[thin,color=black,only marks,mark=o] table[x=x,y expr=\thisrowno{2} < 0 ? \thisrowno{1} : NaN] {data/3d-data/#1/actuators/ac-#2.dat};%
          \else
          \node[above] at (axis cs: 16,32) {#3};%
          \fi%
        }

        \providecommand\makerow[1]{%
          \nextgroupplot%
          \makeplot{output-4x2}{#1}{No offset}%
          \nextgroupplot%
          \makeplot{output-row-4x2}{#1}{$x$-offset}%
          \nextgroupplot%
          \makeplot{output-col-4x2}{#1}{$z$-offset}%
          \nextgroupplot%
          \makeplot{output-row-2x4}{#1}{$2\times4$ equivalent}%
        \pgfmathsetmacro{\mytime}{round((0.2*#1-100))} \node[right] at (axis cs: 32,16) {$t=\pgfmathprintnumber{\mytime}$};
          %\FPeval{\time}{clip(0.2*#1 - 100)}%
          %\node[right] at (axis cs: 32,16) {$t = \time$};%
        }

        \makerow{500}
        \makerow{510}
        \makerow{550}
        \makerow{650}
        \makerow{850}
        \makerow{1000}
      \end{groupplot}
    \end{tikzpicture}
    \caption{Keeping the parameters and number of actuators the same as the unstabilised 3\textsuperscript{rd} column in \cref{fig:3d_interfaces_m} but shifting the actuators (either with an \(x\)-offset to the rows---column 2---or a \(z\)-offset to the columns---columns 3) we see that if we stagger the actuator positions in a grid offset relative to the direction of gravity we can stabilise the previously unstable \(1,1\)-wave.}%
    \label{fig:3d_interfaces_m_shift}
  \end{figure}
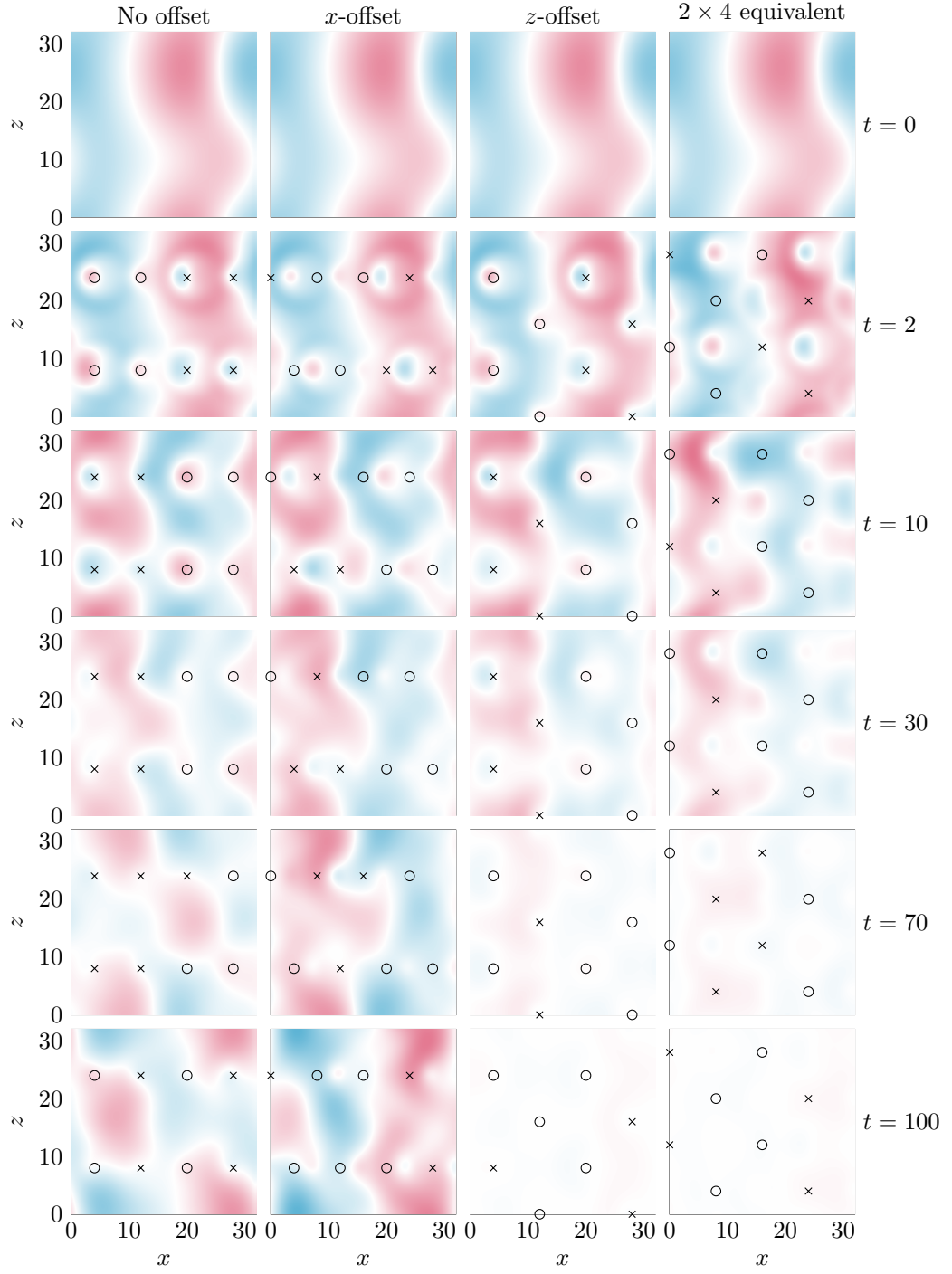

  \begin{figure}[htb]
    \centering
    \begin{tikzpicture}
      \begin{axis}[
        width=0.95\textwidth,
        % height=1.0\textheight,
        axis lines = box,
        xlabel = {\(t\)},
        ylabel = {\(\norm{h-1}\)},
        xmin = -100,
        xmax = 100,
        ymode = log,
        legend pos=south west,
        ]

        \providecommand\makeplot[2]{%
          \addplot [#2]%
            table[x expr=\thisrowno{0}-100, y expr=\thisrowno{1}]%
            {data/3d-data/#1/results.dat};%
        }

        \makeplot{output-2x2}{color=red,solid,}
        \makeplot{output-row-2x2}{color=red,densely dashed,}
        \makeplot{output-col-2x2}{color=red,densely dotted,}

        \makeplot{output-2x4}{color=green,solid,}
        \makeplot{output-row-2x4}{color=green,densely dashed,}
        \makeplot{output-col-2x4}{color=green,densely dotted,}

        \makeplot{output-4x2}{color=blue,solid,}
        \makeplot{output-row-4x2}{color=blue,densely dashed,}
        \makeplot{output-col-4x2}{color=blue,densely dotted,}

        \makeplot{output-4x4}{color=black,solid,}
        \makeplot{output-row-4x4}{color=black,densely dashed,}
        \makeplot{output-col-4x4}{color=black,densely dotted,}

        \legend{
          \(m_x=2\quad m_z=2\) - no offset,
          \(m_x=2\quad m_z=2\) - \(x\)-offset,
          \(m_x=2\quad m_z=2\) - \(z\)-offset,
          \(m_x=2\quad m_z=4\) - no offset,
          \(m_x=2\quad m_z=4\) - \(x\)-offset,
          \(m_x=2\quad m_z=4\) - \(z\)-offset,
          \(m_x=4\quad m_z=2\) - no offset,
          \(m_x=4\quad m_z=2\) - \(x\)-offset,
          \(m_x=4\quad m_z=2\) - \(z\)-offset,
          \(m_x=4\quad m_z=4\) - no offset,
          \(m_x=4\quad m_z=4\) - \(x\)-offset,
          \(m_x=4\quad m_z=4\) - \(z\)-offset,
        }
      \end{axis}
    \end{tikzpicture}
    \caption{Norm of the interfacial perturbations over time for shifted grids of actuators, including the cases shown in \cref{fig:3d_interfaces_m_shift}. We can see that, after the controls are activated at \(t=0\), the inclusion of an offset in the \(z\)-axis makes a very significant difference, especially in the \(m_z = 2\) cases.}%
    \label{fig:3d_lines_m_shift}
  \end{figure}
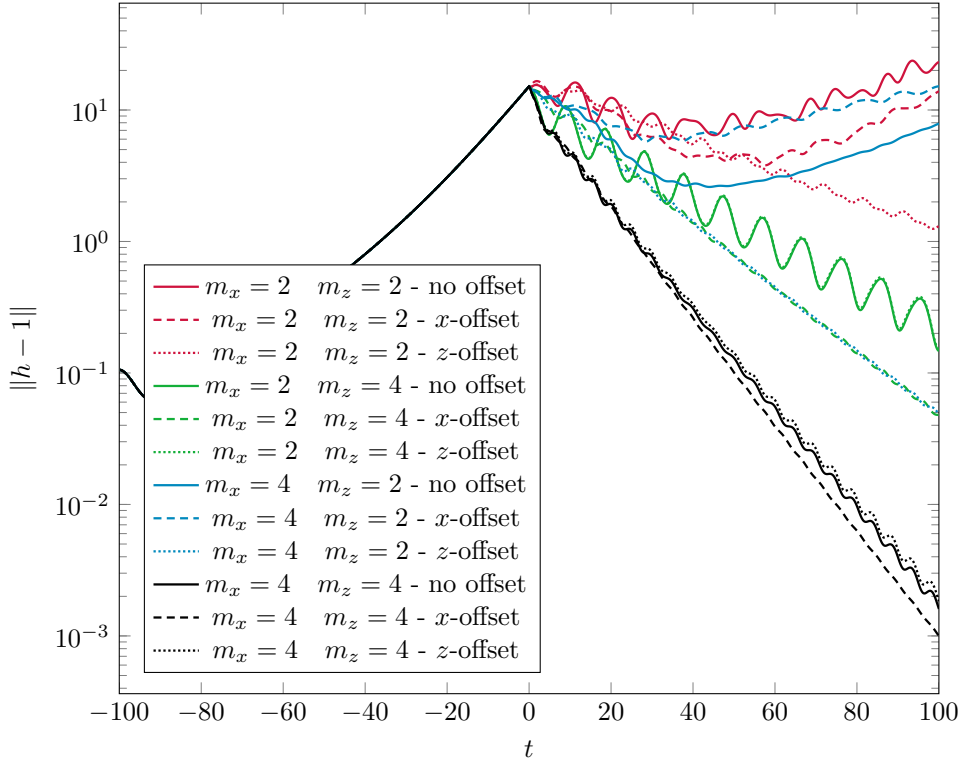

\subsection{Varying the aspect ratio}%
\label{sub:varying_the_aspect_ratio}

In the two-dimensional problem, increasing the domain length $L_x$ permits additional unstable modes to lie within the unstable region of the spectrum (see \cite{holroyd2024linear}).  In three dimensions, the spanwise extent $L_z$ provides a second mechanism for increasing the number of unstable modes. As discussed in \cref{sec:stability_analysis},  increasing $L_z$ causes additional spanwise modes to enter the unstable region.

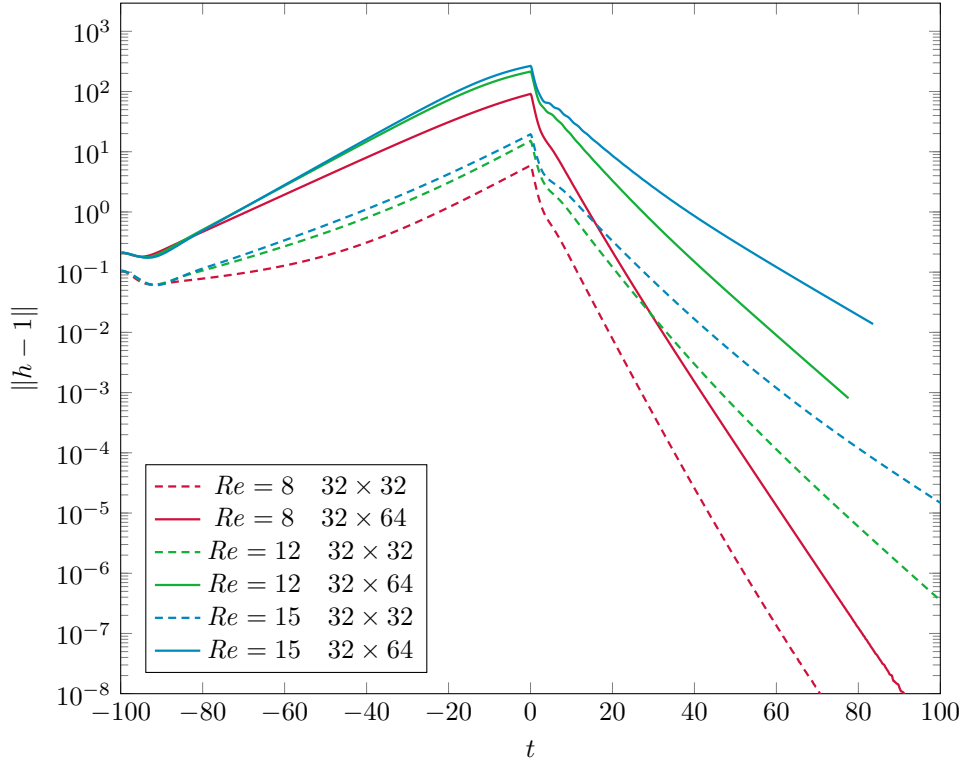
\begin{figure}[htb]
    \centering
    \begin{tikzpicture}
      \begin{axis}[
        width=0.95\textwidth,
        % height=1.0\textheight,
        axis lines = box,
        xlabel = {\(t\)},
        ylabel = {\(\norm{h-1}\)},
        ymin = 0.00000001,
        xmin = -100,
        xmax = 100,
        ymode = log,
        legend pos=south west,
        ]

        \providecommand\makeplot[2]{%
          \addplot [#2]%
            table[x expr=\thisrowno{0}-100, y expr=\thisrowno{1}]%
            {data/3d-data/#1/results.dat};%
        }

        \makeplot{output-8-64x64}{color=red,densely dashed,}
        \makeplot{output-8-64x128}{color=red,solid,}

        \makeplot{output-12-64x64}{color=green,densely dashed,}
        \makeplot{output-12-64x128}{color=green,solid,}

        \makeplot{output-15-64x64}{color=blue,densely dashed,}
        \makeplot{output-15-64x128}{color=blue,solid,}

        \legend{
          \(\Rey = 8 \quad 32\times32\),
          \(\Rey = 8 \quad 32\times64\),
          \(\Rey = 12 \quad 32\times32\),
          \(\Rey = 12 \quad 32\times64\),
          \(\Rey = 15 \quad 32\times32\),
          \(\Rey = 15 \quad 32\times64\)
        }
      \end{axis}
    \end{tikzpicture}
    \caption{Increasing the width of the domain from \(32\) to \(64\) results in larger initial waves developing before the controls are activated. After \(t=0\), simulations with matching values of \(\Rey\) are stabilised at the same rate, due to the proportionally increased number of actuators.}%
    \label{fig:3d_lines_aspect}
  \end{figure}

  In \cref{fig:3d_lines_aspect} we take simulations at three different Reynolds numbers (\(\Rey = 8\), \(12\), \(15\)) and compare the square, \(L_x=L_z=32\), setup to a broader domain with \(L_z = 64\). This introduces an additional \(12\) unstable modes to the
  \(15\) already present in the square problem (the \(8\) of which with a non-zero spanwise component become more unstable). 
  To compensate for these additional unstable modes, the number of actuators in the spanwise direction is increased proportionally from 8 to 16, maintaining a constant actuator spacing across the domain.

  Prior to control activation ($t<0$), the wider domains exhibit larger perturbations due to the presence of additional unstable modes. Once feedback control is activated, however, the perturbations decay at essentially identical rates, as indicated by the approximately parallel trajectories in \cref{fig:3d_lines_aspect}. This suggests that maintaining a fixed actuator density is sufficient to compensate for the increased number of unstable modes introduced by the larger spanwise domain.
  This observation suggests that, at least over the parameter range considered here, control performance depends more strongly on actuator density than on the absolute size of the domain.

\section{Discussion and conclusions}%
\label{sec:4-discussion}

In this study, we showed that the Linear-Quadratic Regulator framework recently developed for two-dimensional falling films can be successfully extended to three-dimensional flows. We demonstrated that, despite the significant additional complexities introduced by the spanwise dimension, the methodology remains robust, with the principal challenge arising from the computation of the gain matrix \(K\).

Through linear stability analysis and direct numerical simulation, we showed that the effects of changing the Reynolds number, capillary number and inclination angle are consistent with those observed in the two-dimensional problem, with the main difference being the introduction of additional unstable spanwise modes.

We further investigated the role of actuator placement, extending ideas previously explored in a related setting within our earlier work \cite{holroyd2025nonlinear}. As in the two-dimensional problem, increasing the number of actuators generally improved stabilisation rates. 
However, the additional spatial degree of freedom also introduced new design considerations. In particular, we showed that modifying the relative spanwise positioning of the actuators can have a substantial effect on performance, increasing damping rates in already stable configurations and, importantly, in some cases even determining whether stabilisation is achieved at all.

From a practitioner's point of view, we believe our findings suggest a natural control design methodology. Linear stability analysis and an understanding of the instability landscape provides a first estimate of the number of actuators required in each spatial direction by identifying (to within the level of accuracy or uncertainty of the theoretical framework) the unstable modes that must be controlled. 
Once this baseline actuator density has been established, actuator placement may be refined to minimise regions in which streamwise or spanwise instabilities can develop. We have explored a shift and rotation of actuator locations, but other configurations can be considered, as discussed by \doublecite{tomlin2019point}, who investigated regular, perturbed, random and quasirandom layouts. 
The final design will inevitably depend on engineering constraints, such as manufacturing tolerances and limitations on the total number of actuators. Importantly, the resulting framework combines analytical predictions with targeted numerical investigations, allowing information that would be difficult to obtain experimentally to be incorporated into the design process prior to implementation.

An interesting question raised by the present results concerns the relative importance of increased model fidelity versus increased physical realism. Recent efforts in thin-film modelling have often focused on deriving increasingly accurate reduced-dimensional models within two-dimensional settings. While such developments remain important, the present work demonstrates that extending existing models to three-dimensional configurations can also yield substantial gains by capturing instability mechanisms that are entirely absent in two dimensions. Determining how best to balance these competing objectives, namely model sophistication and physical dimensionality, remains an open question for future work.

Several important challenges remain.
The periodic domains considered here are both analytically convenient and standard in theoretical studies of falling films. However, experimental systems are naturally characterised by inflow and outflow boundary conditions, and future control frameworks will ultimately need to accommodate such geometries.
Previous studies in both uncontrolled \cite{charogiannis2018experimental} and naively controlled multi-physics settings  \cite{tomlin2020instability} suggest that this is achievable, although the associated computational costs become increasingly demanding.

A second challenge concerns the development of control strategies suitable for real-time deployment, and in particular the assumptions regarding system observations. 
Throughout this work, we have assumed access to detailed information about the interfacial configuration, allowing the feedback law to be evaluated directly. In practical settings, observations are likely to be spatially sparse, available only at discrete time intervals, and contaminated by measurement noise. While related questions were investigated in the two-dimensional setting in \cite{holroyd2025nonlinear}, extending these ideas to three-dimensional flows remains an important direction for future work. The combination of sparse sensing, state estimation and feedback control is likely to play a central role in any experimentally realisable implementation.
Data-driven approaches may provide a natural complement to this effort, either through the construction of surrogate models for rapid state estimation or by enabling efficient reconstruction of unobserved flow quantities from limited measurements.
Data-driven and surrogate modelling approaches may therefore offer a promising route for accelerating controller construction while retaining the key physical mechanisms of the flow. Recent developments in physics-informed learning and reduced-complexity modelling have demonstrated considerable promise in this direction. Experimental validation of such approaches remains an important long-term objective and a natural next step towards practical implementation.

\textbf{Funding.} OAH is grateful for the computing resources supplied by the University of Warwick Scientific Computing Research Technology Platform (SCRTP) and funding from the UK Engineering and Physical Sciences Research Council (EPSRC) grant EP/S022848/1 for the University of Warwick Centre for Doctoral Training in Modelling of Heterogeneous Systems (HetSys CDT). RC and SNG also acknowledge EPSRC Small Grant EP/V051385/1, supporting foundational aspects of this investigation. 

\textbf{Author contributions.} Conceptualisation: Oscar A. Holroyd, Susana N. Gomes, Radu Cimpeanu; methodology: Oscar A. Holroyd, Susana N. Gomes, Radu Cimpeanu; software: Oscar A. Holroyd; formal analysis and investigation: Oscar A. Holroyd; visualisation: Oscar A. Holroyd; writing—original draft: Oscar A. Holroyd; writing—review and editing: Oscar A. Holroyd, Susana N. Gomes, Radu Cimpeanu; supervision: Susana N. Gomes, Radu Cimpeanu; funding acquisition: Susana N. Gomes, Radu Cimpeanu.

\textbf{Data availability.} The computational workflow (including open-source code) for the present work can be found at \url{https://github.com/OaHolroyd/oomph-thin-film-control}. For the purpose of open access, the authors have applied a Creative Commons Attribution (CC-BY) licence to any arising Author Accepted Manuscript version.

% This grouping and its contained commands help with line-breaking and
% justification in the bibliography, which is often problematic.
\begingroup
\setlength{\emergencystretch}{2em}
\hbadness 10000\relax
%\printbibliography % this actually prints the bibliography
\bibliography{references}
\endgroup

\end{document}